\documentclass[11pt,a4paper]{article}

\usepackage{jheppub}
\usepackage{graphicx}
\usepackage{hyperref}

\title{EPOS~LHC : test of collective hadronization with LHC data}

\author[1]{T.~Pierog}

\author[2,3]{Iu.~Karpenko}
\author[4]{J.M.~Katzy}
\author[4]{E.~Yatsenko}
\author[5]{K.~Werner}

\affiliation[1]{Karlsruhe Institute of Technology (KIT), Campus North, Karlsruhe, Germany}
\affiliation[2]{Bogolyubov Institute for theoretical Physics, Kiev, Ukraine}
\affiliation[3]{Frankfurt Institute for Advanced Studies, Frankfurt am Main, Germany}
\affiliation[4]{DESY, Hamburg, Germany}
\affiliation[5]{SUBATECH, University of Nantes - IN2P3/CNRS - EMN, Nantes, France}
\emailAdd{Tanguy.Pierog@kit.edu}

\abstract{
EPOS is a Monte-Carlo event generator for minimum bias hadronic interactions,  
used for both heavy ion interactions and cosmic ray air shower simulations. 
Since the last public release in 2009, the LHC experiments have provided a number of very interesting data sets comprising 
 minimum bias {\it p-p}, {\it p-Pb} and  {\it Pb-Pb} interactions. We describe the changes required to the model 
to reproduce in detail the new data available from LHC and the consequences in the interpretation 
of these data. In particular we discuss the effect of the collective hadronization in {\it p-p} scattering. 
A different parametrization of flow has been introduced in the case of a small volume with high density 
of thermalized matter (core) reached in {\it p-p} compared to  large volume produced in heavy ion collisions. 
Both parametrizations depend only on the geometry and the amount of secondary particles entering in the core
and not on the beam mass or energy. 
The transition between the two flow regimes can be tested with {\it p-Pb} data. 
EPOS~LHC is able to reproduce all minimum bias results for all particles with transverse momentum from $p_t=0$ to a few GeV/c.}

\arxivnumber{1306.0121}

\begin{document}

\maketitle


\section{\label{introduction}Introduction}

Since many years the development of hadronic models able to reproduce with precision
the particle production observed in minimum bias hadronic interactions is a real
challenge. These kind of Monte-Carlo (MC) models are not only important to test 
our knowledge of 
the physical processes involved, but they are also useful in other areas such as to analyze the detector 
acceptance in high energy physics (HEP) experiments or to propagate hadrons in the 
Universe or in the Earth's atmosphere for Astrophysical applications. With the start of
the LHC, a very large data set has become available. Although the models used for
cosmic ray applications were able to predict the general behavior of these 
data~\cite{d'Enterria:2011kw},
 none of them was able to predict all minimum bias data consistently, and 
the models dedicated to HEP such as PYTHIA~\cite{Sjostrand:2006za} 
failed to reproduce accurately distributions involving particles with very low transverse momenta and 
  strange  particle production. 


In the MC generators  commonly used in HEP  
the soft part of the particle production, which dominates the minimum bias 
results,  is dominantly calculated from perturbative QCD and to a small part via diffractive processes. The partons from the 
scattering process are showered and  hadronized following 
the Lund string  or the cluster fragmentation model. 
It was shown in ~\cite{Matinyan:1998ja} 
that at LHC energies  hadronization can not be done with a simple 2 strings 
model without multiple-scattering like for ISR data. 
Many strings can be superimposed and the hadronization can
not be treated like in an empty $e^+e^-$ environment. The need of corrections to the hadronization models 
 has also been acknowledged in \cite{herwigpp, perugiatunes} where it is implemented in form of so called color-reconnection.

\begin{figure}
\includegraphics[height=0.5\columnwidth]{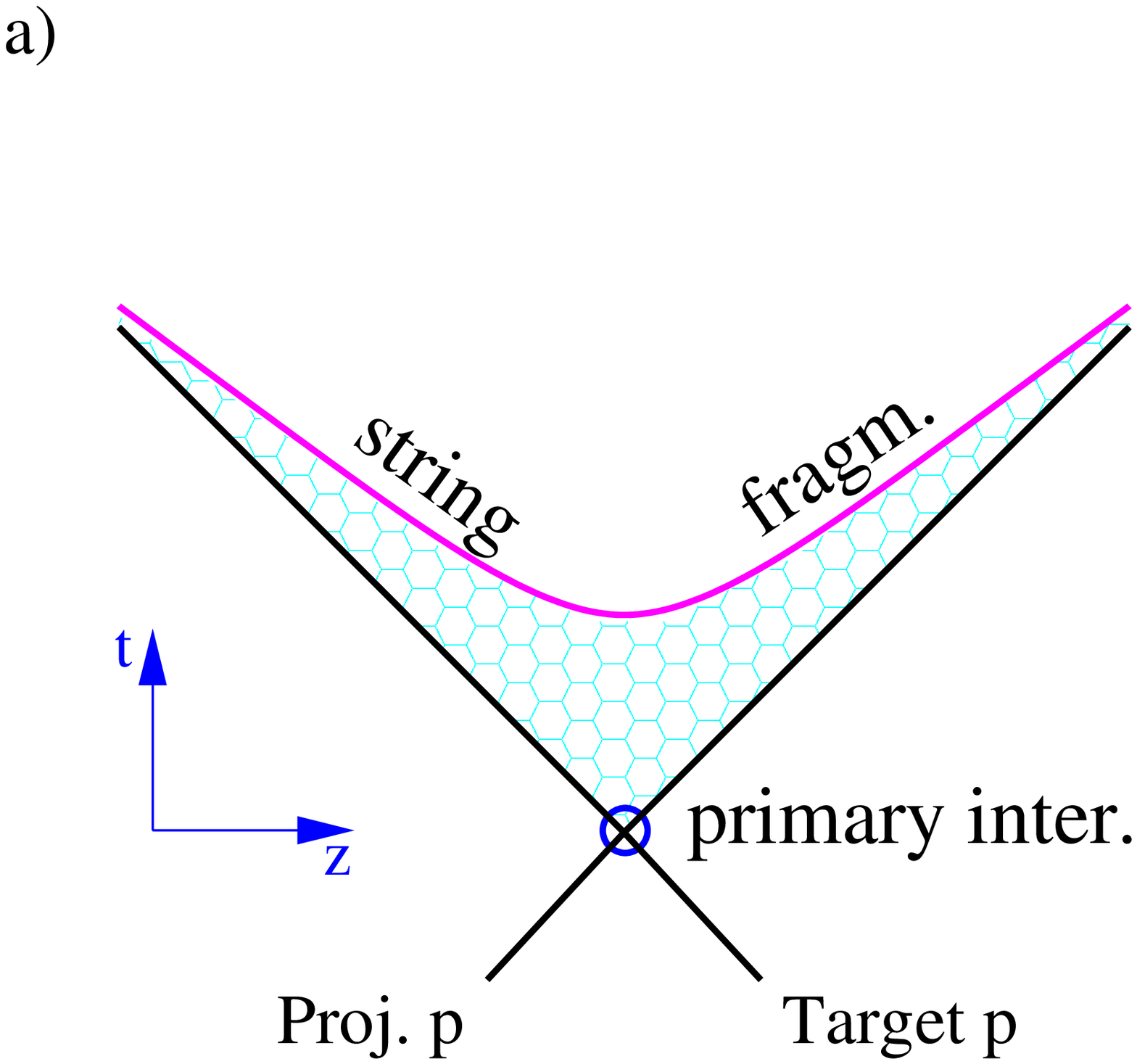}%
\includegraphics[height=0.5\columnwidth]{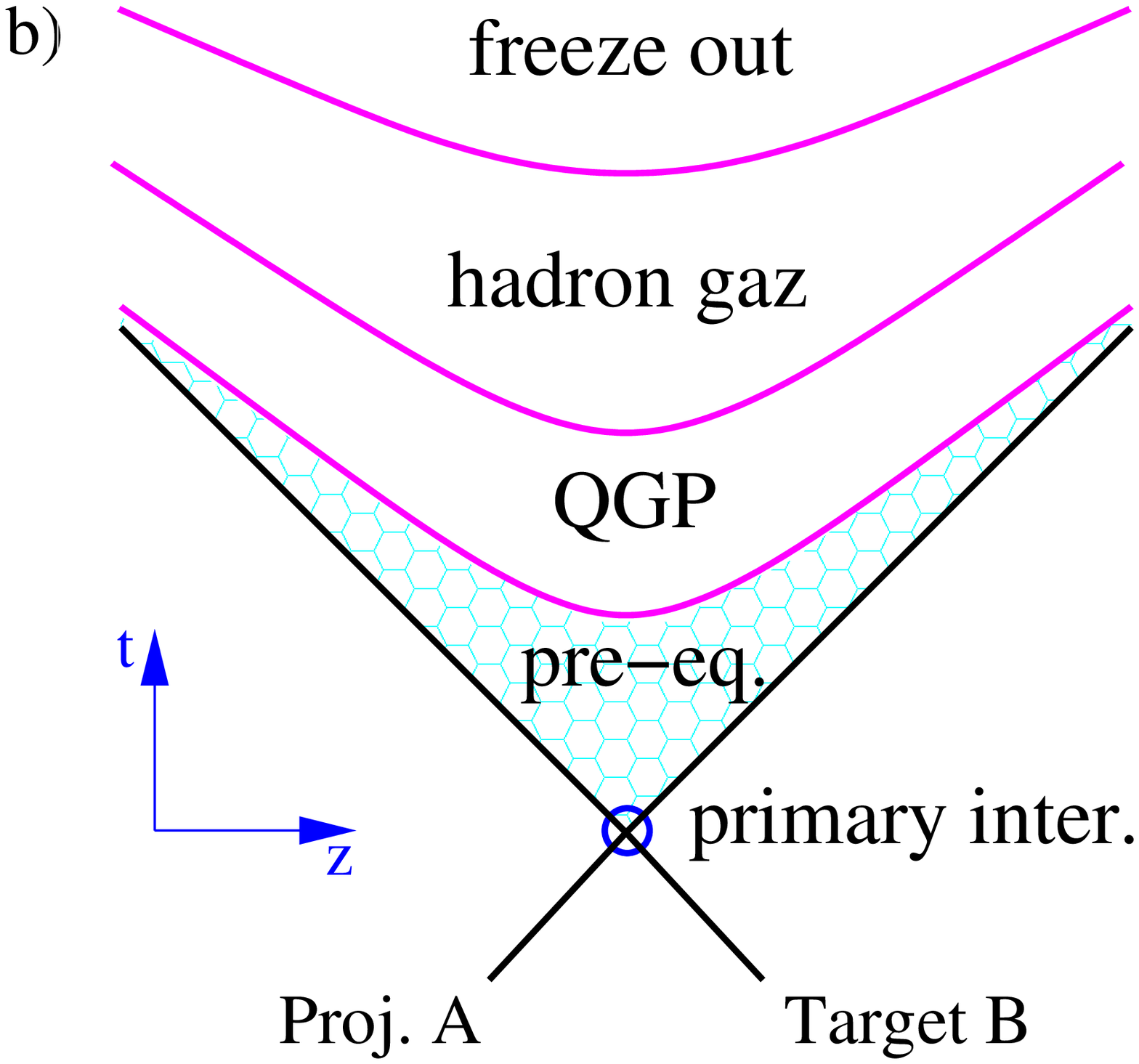}%
\caption{\label{evostd}Space time evolution of the particle production
in an hadronic interaction. An hyperbola (line) represents particles with 
the same proper time. Figure a) is the standard approach for {\it p-p}
scattering while figure b) is a more complete treatment used usually for
HI collision.}
\end{figure}

In figure~\ref{evostd}~a) 
we represent the classical description of a  {\it p-p} scattering in HEP models. 
For LHC {\it p-p} scattering, ``Projectile'' and ``Target'' refer only to two
opposite directions along the beam axis, but the system is naturally
completely symmetric. The aim is to 
 demonstrate that a model based on the complete chain of possible 
hadronic phase as represented in figure~\ref{evostd}~b) commonly used for 
HI collisions can lead to an improved description of minimum bias {\it p-p} data 
at the  LHC. Such a model exists and
is actively developed since about 20 years. EPOS~\cite{Werner:2005jf} 
is based on the Parton-Based Gribov Regge Theory~\cite{Drescher:2000ha} 
developed for NEXUS, which was based on the VENUS model~\cite{Werner:1993uh} 
for soft interactions and  the QGSJET model~\cite{Kalmykov:1997te} for the 
semi-hard scattering.

In this paper we will show how EPOS~1.99~\cite{Pierog:2009zt} 
released in 2009 has been changed to reproduce in detail
LHC data from various experiments.
In section \ref{epos} the basic principle of the model is discussed
before a comparison to data in section \ref{data} with
the updated version EPOS~LHC (v3400). In 
particular  the role of the collective flow which is 
changed
compared to the one used in EPOS 1.99 is demonstrated. Tests with {\it p-Pb} and {\it Pb-Pb'} data 
are shown in section \ref{ion} and finally  the 
difference to the PYTHIA model is discussed in~\ref{pythia}. Only LHC data are 
discussed here
but EPOS~LHC is tuned (with a single parameter set) to reproduced any kind of 
hadronic interactions from
{\it h-A} to {\it A-B} where {\it h} can be $\pi$, {\it K} or {\it p} and {\it A} or {\it B} range from 1 
to 210 nucleons. The energy range is from 40~GeV lab to more than 1000~TeV 
center-of-mass energy (about $10^{21} eV$ lab).\\

The EPOS version EPOS~LHC v3400 presented here differs from EPOS 2.x~\cite{Werner:2010aa}
and EPOS 3.x~\cite{Werner:2013tya} (under development) in that it does not  take advantage of the complete 3D hydro 
calculation followed by the hadronic cascade done in
EPOS~2 or 3, but it is a released version which is freely available for any user~\footnote{available with HepMC interface CRMC at : \url{http://www.auger.de/~rulrich/crmc.html}}.  
The fast covariant approach used in EPOS~1.99 is still used 
but with an improved flow parametrization as described later. 
The main reason to have different versions is that for a {\it Pb-Pb} central event 
EPOS~2 or 3 needs about one hour 
 while EPOS~LHC will generate it in few tens of seconds and EPOS~LHC is not under 
development any more (public stable version). 
As a consequence EPOS~LHC has more parameters (and less predictive power) than
EPOS 2 or 3~\cite{Werner:2011xr,Werner:2012xh,Werner:2013ipa} and should not be used for a 
precise study of $p_t$ distributions or particle correlations in HI collisions, but is
a good alternative model for p-p and p-A minimum bias analysis.

\section{\label{epos}Update of the EPOS~1.99 model}

\subsection{Basic principles of EPOS~1.99}

\begin{figure}[tbh]
\begin{centering}
\hspace*{-0.3cm}\includegraphics[scale=0.37]{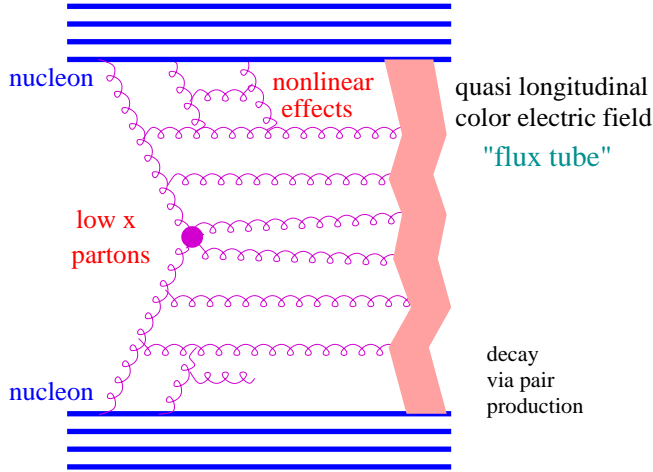}
\par\end{centering}

\caption{Elementary interaction in the EPOS model.\label{cap:Elementary-interaction}}

\end{figure}

Nucleus-nucleus scattering - even proton-proton - amounts to many
elementary collisions happening in parallel. Such an elementary scattering
is the so-called ``parton ladder'' , see figure \ref{cap:Elementary-interaction},
also referred to as cut Pomeron~\cite{Werner:2005jf}. 

A parton ladder represents parton evolutions from the projectile and
the target side towards the center (small $x$). The evolution is
governed by an evolution equation, in the simplest case according
to DGLAP. In the following we will refer to these partons as {}``ladder
partons'', to be distinguished from {}``spectator partons'' to
be discussed later. It has been realized a long time ago that such
a parton ladder may be considered as a quasi-longitudinal color field,
a so-called ``flux tube''~\cite{Werner:2010aa}, conveniently treated as a 
relativistic string. The intermediate gluons are treated as kink singularities
in the language of relativistic strings, providing a transversely
moving portion of the object. This flux tube decays via the production
of quark-antiquark pairs, creating in this way fragments -- which
are identified with hadrons.

The technical details of the
consistent quantum mechanical treatment of the multiple scattering
with the energy sharing between the
parallel scatterings can be found in~\cite{Drescher:2000ha}. 
 Hard scale independent correction factors are 
added to the bare amplitude of the Pomeron to control the rise of the
cross-section at high energy and the multiplicity in HI collisions. The treatment of these 
nonlinear effects at high energy is explained in \cite{Werner:2005jf}.
We don't want to discuss this part of the model
here since very little change has been made compared to EPOS~1.99. Another
article~\cite{epos-lhc} will cover the update of EPOS for the initial part
of the collision (diffraction, string ends and remnants).

\subsection{Collective hadronization in EPOS~1.99}

First of all, it is important to note that the initial conditions for hadronization 
in EPOS are based on strings, not on partons. Here "initial conditions" 
refers to the state of the system after the initial and final state radiation 
of the jets
when partons hadronize in HEP models, and before possible parton or hadron
rescattering like in HI collisions (final state interactions).
As explained in previous section, the initial scatterings
lead to the formation of strings, which break into segments, which
are usually identified with hadrons. Then one considers the situation 
at an early
proper time $\tau_{0}$, long before the hadrons are formed: one distinguishes
between string segments in dense areas (more than some critical density
$\rho_{0}$ segments per unit volume), from those in low density areas.
The high density areas are referred to as core, the low density areas
as corona~\cite{Werner:2007bf}. The corona is important for certain aspects
like the centrality dependence of all observables in HI collisions. Here it
will correspond to unmodified string fragmentation like in usual HEP models
and will dominate at large rapidity and in low multiplicity events.
In this section we will focus on the core part which is unique in EPOS
and provide interesting effects not taken into account in other HEP models
(which are all ``corona''-like). 

Based on the four-momenta
of the string segments which constitute the core, a matrix in
$(x,y,\eta)$ of the segment density is formed. The core is made of different 
clusters
in each $\eta$ bin to keep the local energy density distribution and each
cluster is hadronized via a microcanonical procedure with an additional
longitudinal and radial flow exactly as described 
in~\cite{Werner:2007bf}. The whole procedure perfectly conserves energy,
momentum, and flavors. The free parameters used in this process like strangeness
or baryon production correction factor and energy density at freeze-out can
be fixed using HI data on particle production. The mass $M$ of each cluster is 
defined as

\begin{equation} \label{eq:mass}
M = \sqrt{(\sum_{i}E_{i})^2-(\sum_{i}\overrightarrow{P_{i}})^2 }
\end{equation}

where $i$ is the index of all segments forming the cluster and 
$(E_{i},\overrightarrow{P_{i}})$ the four-momentum vector of a segment.

\begin{figure}[hptb]
\begin{centering}
\includegraphics[height=0.6\columnwidth]{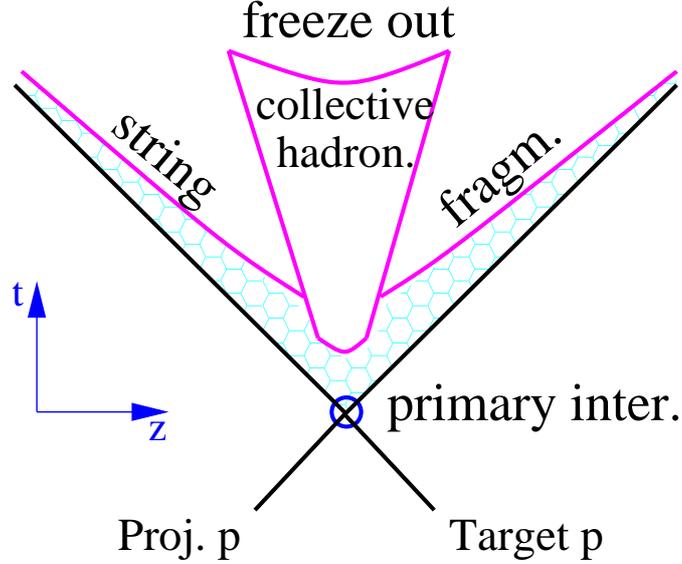}%
\caption{\label{evoepo}Schematic view of the space time evolution of 
the particle production in an hadronic interaction in EPOS~1.99 or EPOS~LHC. 
An hyperbola (line) represents particles with the same proper time. The same
treatment is used for {\it p-p} or {\it A-B} but the collective hadronization, which
can be local, is simplified compared to the full HI picture (done in 
EPOS 2 or 3).}
\end{centering}
\end{figure}

Event-by-event a part of the string segments hadronizes normally (corona) 
and a part is 
used to create a core with a collective hadronization as represented on
figure~\ref{evoepo}. The core appears only if the local density of string 
segments is high enough. This limit is of course easily reached in case
of central HI collisions at RHIC or LHC (or even SPS) because of the 
large number of pairs of nucleons suffering an inelastic interaction.

\begin{figure}[hptb]
\begin{centering}
\includegraphics[height=0.6\columnwidth,angle=-90]{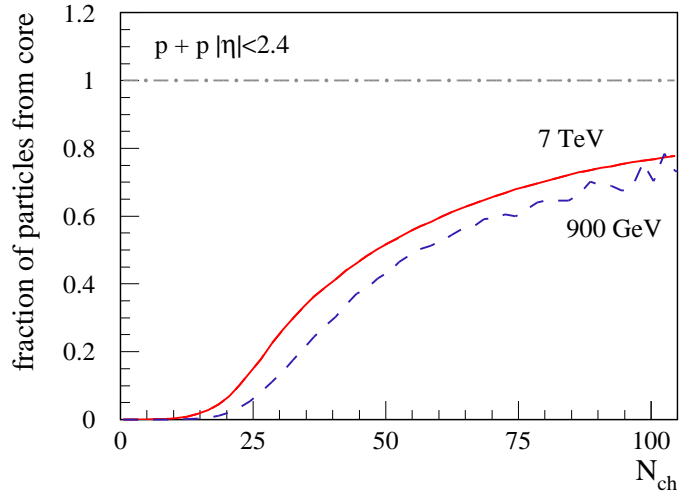}%
\caption{\label{fcore}Fraction of charged particles with $|\eta|<2.4$
coming from the core as a function of the total number of charged particles
with $|\eta|<2.4$. Solid line is used for simulation with EPOS at 7~TeV 
and dashed line for 900~GeV {\it p-p} scattering.}
\end{centering}
\end{figure}

But in fact the multiple scattering of partons for a given pair of nucleons
can be enough to create many strings which will overlap since the distance
between partons is very small. At 7~TeV in {\it p-p} it is easily possible to
 produce more than 5 flux tubes leading to the production of much more than 10
strings very close to each other in the transverse plane and overlapping 
around $\eta=0$. Since a string produce more than a minimum given number of particles, a large
number of strings implies a large multiplicity. So plotting the 
fraction of final particles produced by core decay as a function of the
multiplicity of charged particles with $|\eta|<2.4$ as shown in figure~\ref{fcore} we can
notice that even for the average multiplicity at 7~TeV (solid line), which is 
about 30 (with 4 to 6 strings in average), 
about 30\% of the particles are coming from the core. The rest is produced 
directly by the string fragmentation in corona region where string segments do 
not overlap. At 900~GeV 
(dashed line), at the average multiplicity (about 15), this fraction is close 
to 0. But for the same number of particles, about the same ratio is reached at
 both energies.

Of course not all particles are completely absorbed in the dense area. We 
define a parameter $p_{\rm t}^{\rm cut}$ above which a particle will simply lose
part of its momentum in the core but will survive as an independent particle
produced by a string (typically high $p_{\rm t}$ particles from jets). Soft
particles with $p_{\rm t}<p_{\rm t}^{\rm cut}$ are completely absorbed and form the
core. The energy loss depends on the system size and follow formula 
from~\cite{Baier:1996kr,Peigne:2008ns}.

\subsection{Collective Flow}\label{collflow}

\subsubsection{Flow definitions in EPOS~1.99}

Since string segments show a Bjorken-fluid like behavior, and clusters are formed from
these segments, clusters are considered to be collectively expanding : 
\begin{itemize}
\item some transverse expansion driven by the maximal radial 
rapidity $y_{\rm rad}$ if the total mass of the 
core $M_{\rm core}=\sum_{\rm clusters}M$ is larger than $M_{min}=3~GeV/c^2$,
\item Bjorken-like expansion in longitudinal direction driven by the 
maximal longitudinal rapidity $y_{\rm long}$ for each individual 
cluster whose mass $M$ is larger than the minimum mass $M_{min}$.
\end{itemize}
In~\cite{Werner:2007bf}, it is assumed that the clusters hadronize
at some given energy density $\varepsilon_{\rm hadr}$, having acquired at that 
moment a collective radial flow, with a linear radial rapidity profile from
inside to outside, characterized by $y_{\rm rad}$.
In addition, an azimuthal asymmetry was imposed by multiplying the $x$ and $y$
component of the flow four-vector-velocity with $1+\min(\epsilon,f_{\rm ecc})$ and
$1-min(\epsilon,f_{\rm ecc})$, where $\epsilon$ is the initial spacial 
eccentricity, $\epsilon=\langle (y^2 - x ^2) / (y^2 + x ^2) \rangle$, and
$f_{\rm ecc}=0.5$ a parameter. By imposing radial flow, the cluster mass had to be 
rescaled as

\begin{equation} \label{eq:radmass}
M \rightarrow M \times 0.5 y^{2}_{\rm rad}/(y_{\rm rad}\sinh y_{\rm rad} - \cosh y_{\rm rad} + 1),
\end{equation}

in order to conserve energy and for the longitudinal flow we have in addition

\begin{equation} \label{eq:longmass}
M \rightarrow M \times y_{\rm long}/\sinh y_{\rm long}.
\end{equation}

\subsubsection{New features in EPOS~LHC}

As a consequence of the rescaling due to collective flows, and in particular
the radial flow, the number of secondary particles produced by the clusters
is reduced. In case of a consistent treatment of cross-section and particle
production like in EPOS, this property is needed in the case
of HI collisions where less particles are observed than produced
by the model without final state interactions. And indeed a proper hydro
treatment like in EPOS~2 or 3 requires a large multiplicity in the initial state
to finish with the correct multiplicity after a long evolution of the 
large volume of the core. We will call it the
nuclear AA flow, characterized by the maximal radial 
rapidity $y^{\rm AA}_{\rm rad}$.

But in the case of light system, like {\it p-p}, using EPOS~2 or 3 with a realistic
treatment of the hydrodynamical evolution with proper hadronization such an 
effect was not observed~\cite{Werner:2010ny,Werner:2013ipa,Werner:2013tya}. 
In that case the large flow
comes from the quick expansion of the very small volume of the core.
As a consequence, in EPOS~LHC we 
introduced a 
different type of radial flow in case of very dense system in a small volume
(where the critical
energy density is reached because of multiple scattering between partons in
a single pair of nucleons like in {\it p-p}). For this pp flow, characterized by the 
maximal radial rapidity $y^{\rm pp}_{\rm rad}$, the mass of the cluster $M$
is not changed before hadronization (multiplicity is conserved) but the
energy conservation is imposed by a simple rescaling of the total momentum $P$
(larger $p_{\rm t}$ are compensated by smaller $p_{\rm z}$) after the radial boost. Of course a smooth transition is needed between the two kinds of system and the
transition is observed in {\it p-A} interactions.

In EPOS~1.99, $y_{\rm rad}$ was parametrized as function of the system energy and size as  $y_{\rm rad}=y^{\rm mx}_{\rm rad}+y^{\rm mi}_{\rm rad}\log (1.+\sqrt{s/N_{\rm pair}})$, where $N_{\rm pair}$ is the number of possibly interacting pairs of nucleons and $y^{\rm mx}_{\rm rad}$ and $y^{\rm mi}_{\rm rad}$ are parameters.
While the evolution with $N_{\rm pair}$ was safe and easy to test with HI
data at SPS and RHIC, the evolution with energy especially for $N_{\rm pair}=1$ 
({\it p-p}) could lead to wrong extrapolation at high energy. 

In the paper on identified particle spectra from CMS~\cite{Chatrchyan:2012qb},
data show that the increase of the $\langle p_{\rm t} \rangle$ as a function
of the multiplicity doesn't depend on the center-of-mass energy but increase
with multiplicity. This effect being directly link to the radial flow intensity
in our approach~\cite{Werner:2010ny} it is natural to parametrized all flows
as a function of the total mass $M_{\rm core}$ which is directly link to the final multiplicity. We get for the longitudinal flow
\begin{equation}\label{eq:longflow}
y_{\rm long}=y^{\rm mx}_{\rm long}\cdot\log(\exp (\frac{y^{\rm mi}_{\rm long}}{y^{\rm mx}_{\rm long}} )+ \frac{M_{\rm core}}{M_{min}}),
\end{equation}
the AA radial flow
\begin{equation}\label{eq:AAflow}
y^{\rm AA}_{\rm rad}=y^{\rm mx}_{\rm rad}\cdot\log(\frac{M_{\rm core}}{M_{min}}),
\end{equation}
and the pp radial flow
\begin{equation}\label{eq:ppflow}
y^{\rm pp}_{\rm rad}=y^{\rm px}_{\rm rad}\cdot\log(\frac{M_{\rm core}}{M_{min}})
\end{equation}
where  $y^{\rm mx}_{\rm rad}$ is the
parameter fixing the radial flow and the multiplicity in HI 
and $y^{\rm px}_{\rm rad}$ determine
the evolution of $\langle p_{\rm t} \rangle$ as a function of the multiplicity
in {\it p-p}.
The longitudinal flow being weaker, $y^{\rm mx}_{\rm long}$ and $y^{\rm mi}_{\rm long}$ are parameters which are used to do a fine tuning of the multiplicity in
both {\it p-p} ($y^{\rm mi}_{\rm long}$) and HI interactions ($y^{\rm mx}_{\rm long}$).
 $M_{0}=3 GeV/c^2$ is the minimum mass to have flow.

These definitions are good if we have pure AA flow (slow expansion of a large
volume) or pure pp flow (fast expansion of a small volume). For peripheral nuclear collision or 
{\it p-A} interactions we have to decide whether $y^{\rm pp}_{\rm rad}$ (high density 
of string segments coming from a single nucleon pair like in {\it p-p}) or 
$y^{\rm AA}_{\rm rad}$  (high density of string segments coming from many 
different nucleon pairs like in central HI) should be used. So for pp flow 
parametrization one can defined 
\begin{equation} \label{eq:Mflow}
M_{\rm pp}=\min\left( 1, f_{\rm pp}\cdot\frac{N^{pp}_{max}}{N_{tot}}\right)\cdot M_{\rm core},
\end{equation}
where $N^{pp}_{max}$ is the maximum number of segments used in core coming from
a unique nucleon-nucleon pair and $N_{tot}$ is the total number of segments 
forming the core. So $N^{pp}_{max}/N_{tot}$ is simply the fraction of segments
coming from the pair contributing the most to the core. $f_{\rm pp}=1.3$ is a
parameter which can be tune in order to have the same $\langle p_{\rm t} \rangle$
 as a function of multiplicity in {\it p-p} or {\it p-A} at low multiplicity (no data 
needed). In other words $M_{\rm pp}$ is simply the mass of the part of the core 
coming from the pair of nucleon with the largest multiple scattering. As a
consequence eq.~\ref{eq:ppflow} can be rewritten as
\begin{equation}\label{flowpp}
y^{\rm pp}_{\rm rad}=y^{\rm px}_{\rm rad}\cdot F_{\rm pp} \cdot \log(\frac{M_{\rm pp}}{M_{min}}),
\end{equation}
where $F_{\rm pp}=\min(1,2 \langle N^{pp} \rangle / N^{pp}_{max})^2$, with 
$\langle N^{pp} \rangle$ being the average number of segments going to core per
participating pair of nucleons, is a normalization factor going to 0 in 
case of HI collisions. As expected $F_{\rm pp}=1$ and 
$M_{\rm pp}=M_{\rm core}$ in case of {\it p-p} scattering. Using eq.~\ref{eq:AAflow}, we have 2 independent definitions for 
$y^{\rm pp}_{\rm rad}$  and $y^{\rm AA}_{\rm rad}$. Since $y^{\rm pp}_{\rm rad} 
> y^{\rm AA}_{\rm rad}$ for small system (small number of nucleon pairs) and
$y^{\rm pp}_{\rm rad} < y^{\rm AA}_{\rm rad}$ for large system (when  
$\langle N^{pp} \rangle \ll N^{pp}_{max}$: $F_{\rm pp}\rightarrow 0$), we simply
use the pp flow when $y^{\rm pp}_{\rm rad} > y^{\rm AA}_{\rm rad}$ and the AA flow
otherwise.

Since the flows depend only
on  $M_{\rm core}$ all parameters can be fixed using LHC data only ({\it p-p} and 
{\it Pb-Pb}) but the results are checked with RHIC and SPS data.

\section{\label{data}Comparison with LHC data}

In this section we will see how LHC data can be described by EPOS~LHC.

\subsection{Cross-section}

\begin{figure}[hptb]
\begin{centering}
\includegraphics[height=0.6\columnwidth,angle=-90]{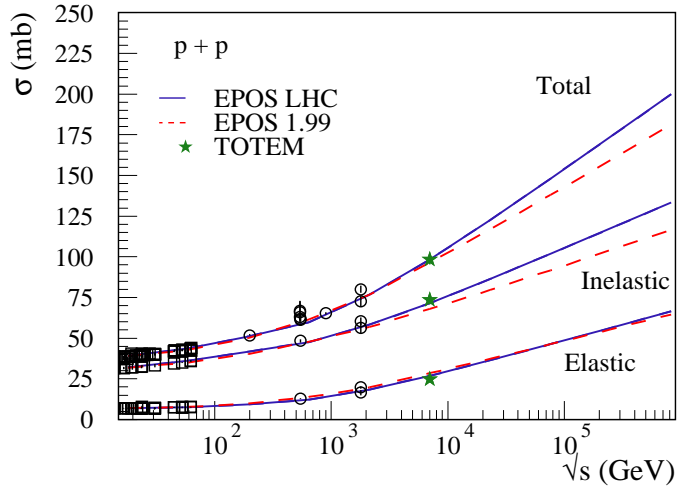}%
\caption{\label{sig}Total, inelastic and elastic {\it p-p} cross section calculated with EPOS~LHC (solid line) and EPOS~1.99 (dashed line). Points are data from~\protect\cite{PDG98} and the stars are the LHC measurements by the TOTEM experiment~\protect\cite{Csorgo:2012dm}.}
\end{centering}
\end{figure}

The most fundamental parameters of the EPOS model are fixed by comparing the 
cross-sections calculated from the single scattering amplitude with the measured
data. Thanks to the TOTEM experiment~\cite{Csorgo:2012dm}, total, inelastic and elastic cross-sections
are known now with a high precision and can be used to constrain the model
at high energy. As
shown on figure~\ref{sig}, after retuning of the parameters fixing the non-linear
effects in EPOS~LHC, both total, inelastic and elastic cross-sections can be well
reproduced (solid line).

Compared to EPOS~1.99 (dashed line), it corresponds to an increase of the inelastic cross
section due to a larger amplitude of the parton ladder. As a consequence the
multiplicity predicted by the model should increase at LHC.

\subsection{Particle production}\label{particle}

\begin{figure}[hptb]
\begin{minipage}[t]{0.48\linewidth}
\begin{centering}
\includegraphics[height=1.\columnwidth,angle=-90]{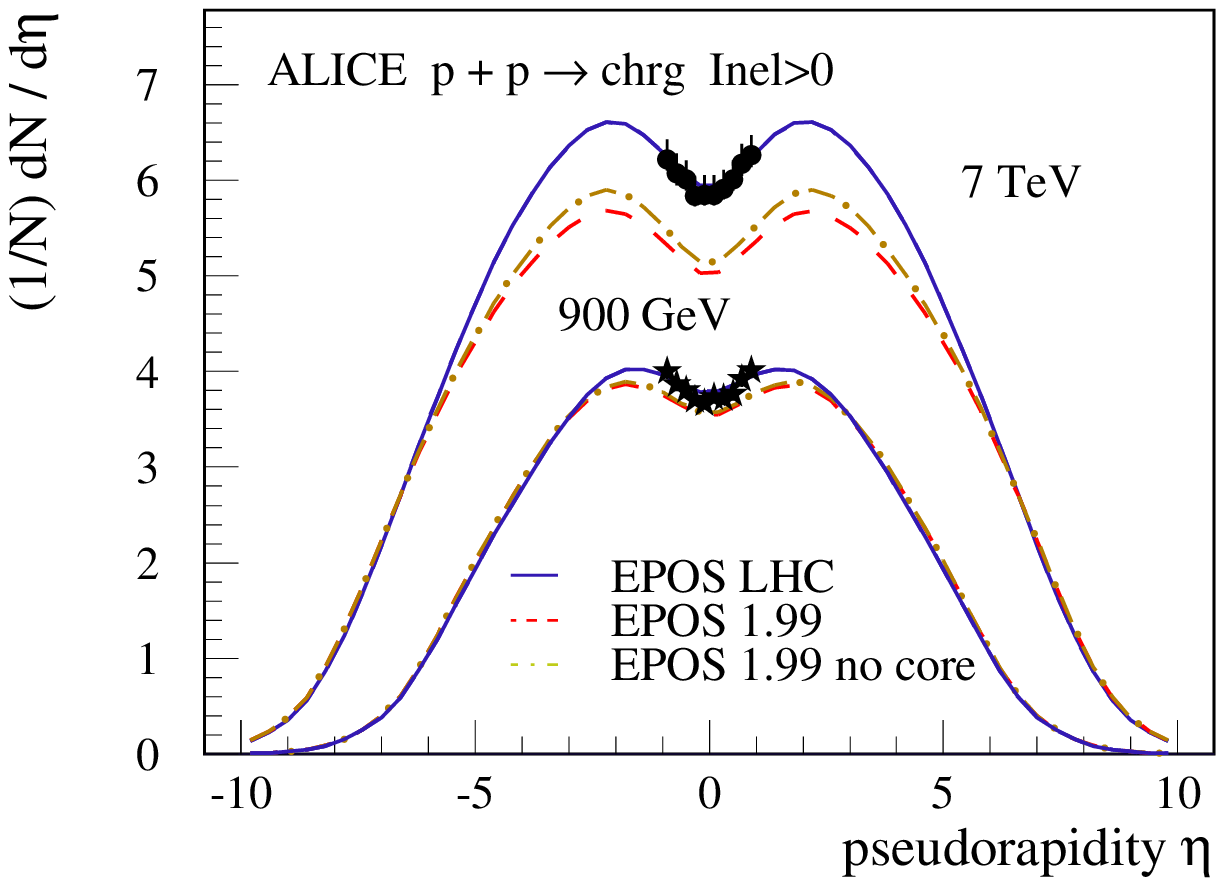}%
\caption{\label{eta}Pseudorapidity distribution $dN/d\eta$ of charged particles for events with at least one charged particle with $|\eta|<1$ for {\it p-p} interactions at 900~GeV and 7~TeV. Simulations with EPOS~LHC (solid line), EPOS~1.99 (dashed line), and EPOS~1.99 without core production (dash-dotted line) are compared to data points from ALICE experiment~\protect\cite{Aamodt:2010pp}.}
\end{centering}
\end{minipage}
\hfill
\begin{minipage}[t]{0.48\linewidth}
\begin{centering}
\includegraphics[height=1.\columnwidth,angle=-90]{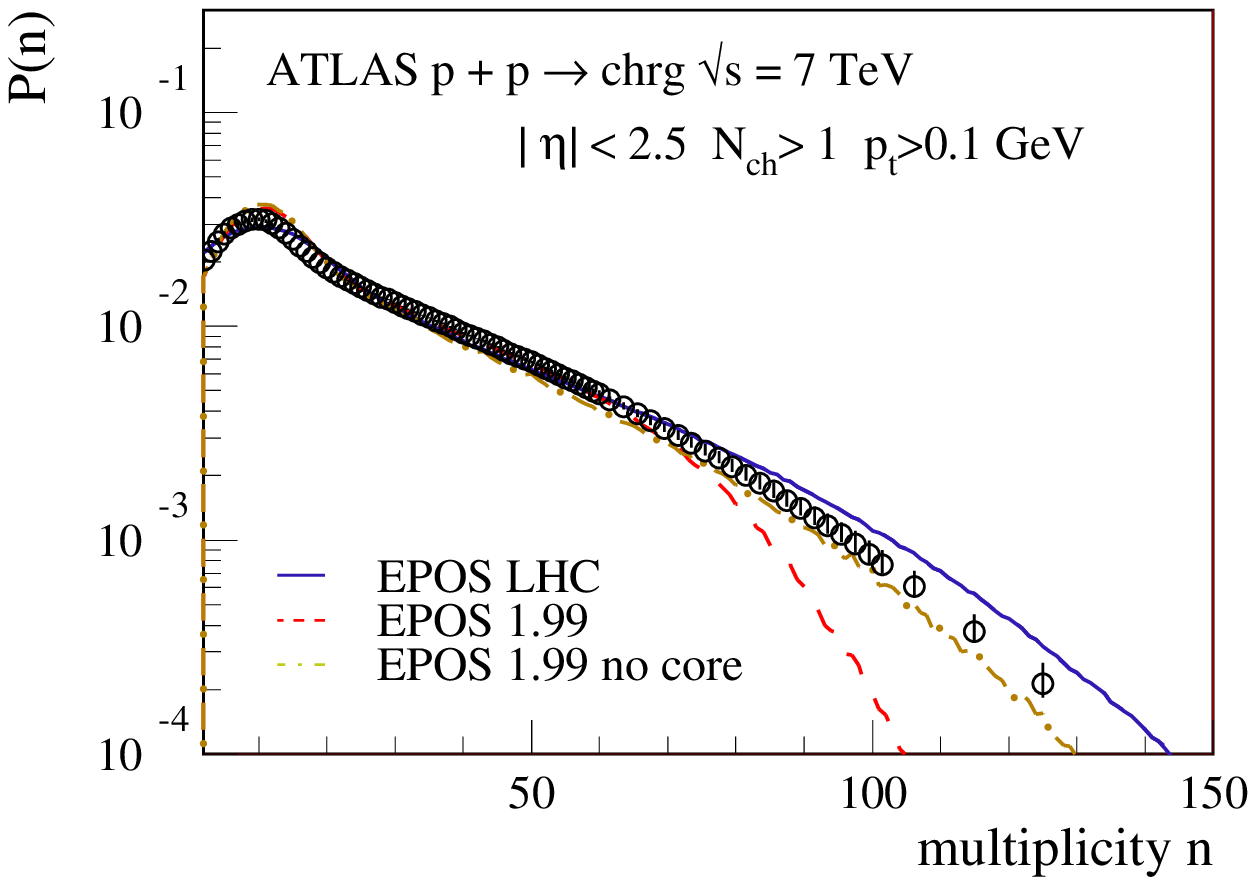}%
\caption{\label{mul}Multiplicity distribution of charged particles with 
$p_t>$200~MeV and $|\eta|<2.5$ for {\it p-p} interactions at 7~TeV. Simulations 
with EPOS~LHC (solid line), EPOS~1.99 (dashed line), and EPOS~1.99 without core production (dash-dotted line) are compared to data points from ATLAS collaboration~\protect\cite{Aad:2010ac}}
\end{centering}
\end{minipage}
\end{figure}

As we can see on figure~\ref{eta} on the comparison of EPOS with pseudorapidity
distribution of charged particles from ALICE data~\cite{Aamodt:2010pp}, 
the mean multiplicity is indeed larger in EPOS~LHC (solid line) compared to
EPOS~1.99 (dashed line). From the simulations with EPOS~1.99 without core
formation (dash-dotted line), we can check that this is not due to the 
corrected flow. The increase in multiplicity is a direct consequence 
of the consistent treatment~\cite{Drescher:2000ha}
of the cross-section and the particle production in EPOS framework. By adjusting
the parameters to get the correct cross-section, we obtain naturally the 
correct multiplicity.

Looking at the corresponding multiplicity distribution at 7~TeV from ALICE 
experiment~\cite{Aad:2010ac} on figure~\ref{mul}, we can observe the effect
of the corrected flow on the tail of the distribution. In EPOS~1.99 (dashed 
line) for the
events with a large multiplicity, the flow effect was strong and was reducing
the total number of particles suppressing events with large multiplicities.
Without core formation in EPOS~1.99 (dash-dotted line), the results were 
already reasonable and now in EPOS~LHC the tail is well reproduced event with 
core formation (solid line).

\begin{figure}[hptb]
\begin{centering}
\includegraphics[height=0.6\columnwidth,angle=-90]{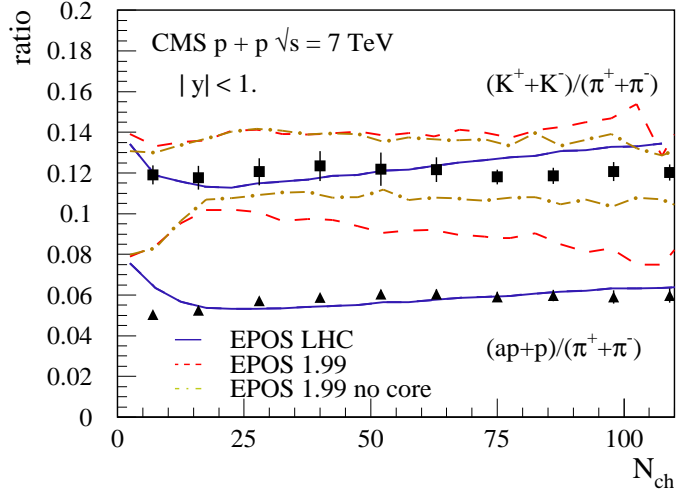}%
\caption{\label{rid}Ratio of particle yield as a function of charged
particle multiplicity for $|y|<1$ for non-single diffractive (NSD) {\it p-p} scattering at 7~TeV. Simulations are done with EPOS~LHC (solid line), EPOS~1.99 (dashed line) and EPOS~1.99 without core production (dash-dotted line). Points are data from CMS experiment~\protect\cite{Chatrchyan:2012qb}.}
\end{centering}
\end{figure}

Thanks to CMS data on identified charged particle ratios at 
mid-rapidity~\cite{Chatrchyan:2012qb} we could identify a problem in EPOS~1.99 
concerning the production of baryon-antibaryon pair (and strangeness) in string 
fragmentation
(dashed line in figure~\ref{rid}) which were artificially increased at high 
energy (by a factor of 2 for diquark production !). After 
correction, and using the same parameters as in $e^+e^-$ string fragmentation,
the data are now well reproduced for both Kaon and (anti)proton production by
EPOS~LHC (solid line in figure~\ref{rid}). The effect of the core formation on
these types of particles is small as we can see for the Kaon ratio 
comparing EPOS~1.99 with and without core (dashed-dotted line in 
figure~\ref{rid}). For the proton ratio using EPOS~1.99 we can clearly see the 
transition 
from a particle production dominated by the (wrong) string fragmentation at 
low N$_{\rm ch}$ to a particle production dominated by cluster decay at
large N$_{\rm ch}$ were the proton to pion ratio is correct. 

With correct 
string fragmentation parameters like in EPOS~LHC, the transition from pure 
strings to clusters is difficult to observe with these type of ``light quark'' 
particles. We will see in the next section that the effect of the final
state interaction is much larger for multi-strange baryons.

\subsection{Final state interaction}

We just reported in section~\ref{particle} that, when everything is treated
correctly, the effect of a possible statistical decay phase with a radial 
flow due to parton or hadron reinteraction after the initial state interaction
is difficult to observe in (light) particle multiplicity with a long life time.
We will see in this section that specific observables show clear indications
that indeed final state interaction effects clearly help to reproduce {\it p-p}
particle production.

\subsubsection{Particle ratio}

\begin{figure}[hptb]
\begin{minipage}[t]{0.48\linewidth}
\begin{centering}
\includegraphics[height=1.\columnwidth,angle=-90]{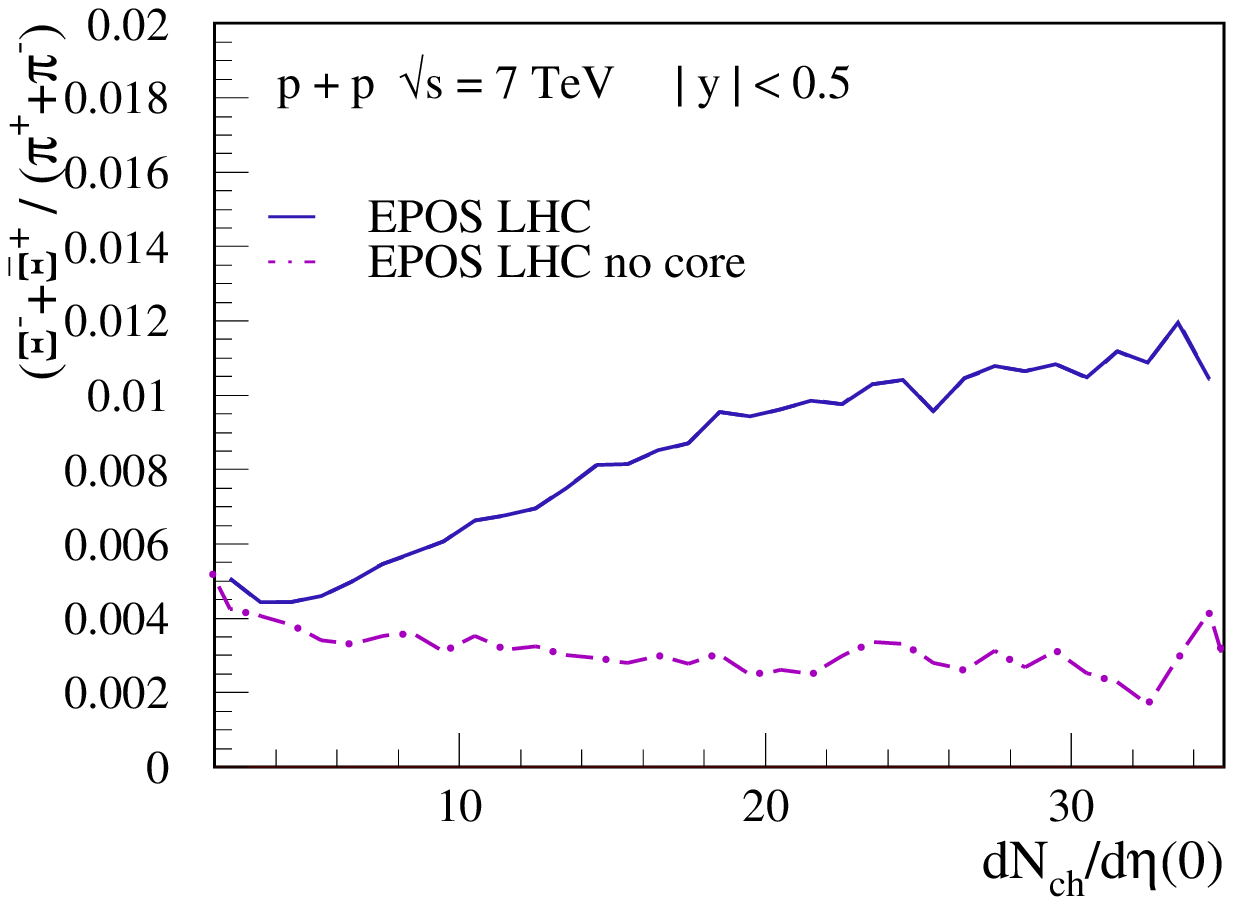}%
\caption{\label{ratioxi}Ratio $\Xi/\pi$ as a function of particle density
at midrapidity from EPOS~LHC simulations with core (solid line) and without 
core production (dash-dotted line).}
\end{centering}
\end{minipage}
\hfill
\begin{minipage}[t]{0.48\linewidth}
\begin{centering}
\includegraphics[height=1.\columnwidth,angle=-90]{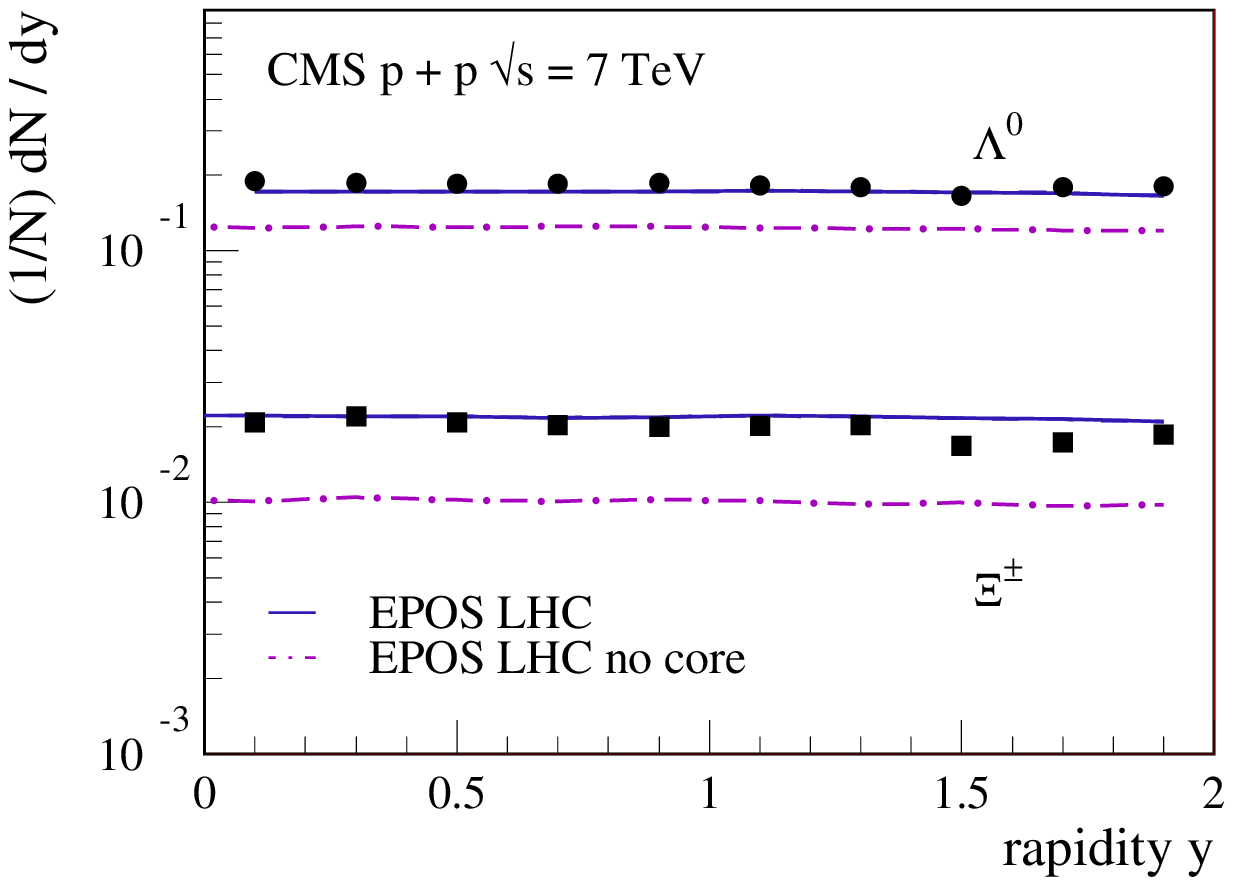}%
\caption{\label{rapstr}Strange baryon yield as a function of rapidity
for non-single diffractive (NSD) {\it p-p} scattering at 7~TeV. Simulations are done with EPOS~LHC with core (solid line) and without core production (dash-dotted line). Points are data from CMS experiment~\protect\cite{Khachatryan:2011tm}.}
\end{centering}
\end{minipage}
\end{figure}

One of the effect of the reinteraction is that the particle production is not
only coming from string fragmentation (where all the parameters are suppose 
to be
fixed by data on $e^+e^-$ particle production) but that part of the particles 
are coming from a phase were particles are produced statistically. In such an
hadronization process corresponding to an hadron gas in equilibrium, strangeness
production is not suppressed. In practice the equilibrium is not necessarily 
reached and some suppression (as free parameter) can be introduced but it is 
shown in~\cite{Becattini:2000jw} that the strangeness production is much larger 
(about a factor of 2) in HI collision (where the parameters for statistical 
hadronization are fixed)
 than in $e^+-e^-$ interactions (where string fragmentation parameters are 
fixed). In \cite{Pop:2012ug} for instance, the hyperon to meson ratio
 is used as a possible proof of a mini-Quark-Gluon-Plasma. In EPOS~LHC, we can
clearly see the transition from a pure string fragmentation to a statistical
dominated hadronization looking at the evolution of the multi-strange baryon
 to pion ratio as a function of the multiplicity at mid-rapidity as shown
figure~\ref{ratioxi}. Without core formation (dash-dotted line), there is no
strong evolution while with core formation (solid line) the ratio increases
almost linearly with the plateau height. The effect would be even larger with
$\Omega$ baryon.

Comparing, figure~\ref{rapstr}, $\Lambda^0+\overline{\Lambda}^0$ and 
$\Xi^- + \overline{\Xi}^+$  rapidity distribution as measured by CMS 
experiment~\cite{Khachatryan:2011tm} for NSD events with EPOS~LHC simulation 
with (solid line) core formation, we observe a good agreement. While without 
(dash-dotted line) core formation the average $\Xi$ production is a factor of
2 lower.

\subsubsection{Transverse momentum}

The second main effect of the collective phase is the generation of a 
collective flow as described in \ref{collflow}. After cluster decay, 
a random longitudinal Lorentz boost whose maximal value is given by 
eq.~\ref{eq:longflow} and a radial Lorentz boost, whose maximal value is
given by eq.~\ref{flowpp} and whose phase depends on cluster geometry, 
are applied to each particle. The effect is better observed in the evolution
of the mean $p_{\rm t}$ as a function of the number of particles at mid-rapidity.

\begin{figure}[hptb]
\begin{minipage}[t]{0.48\linewidth}
\begin{centering}
\includegraphics[height=1.\columnwidth,angle=-90]{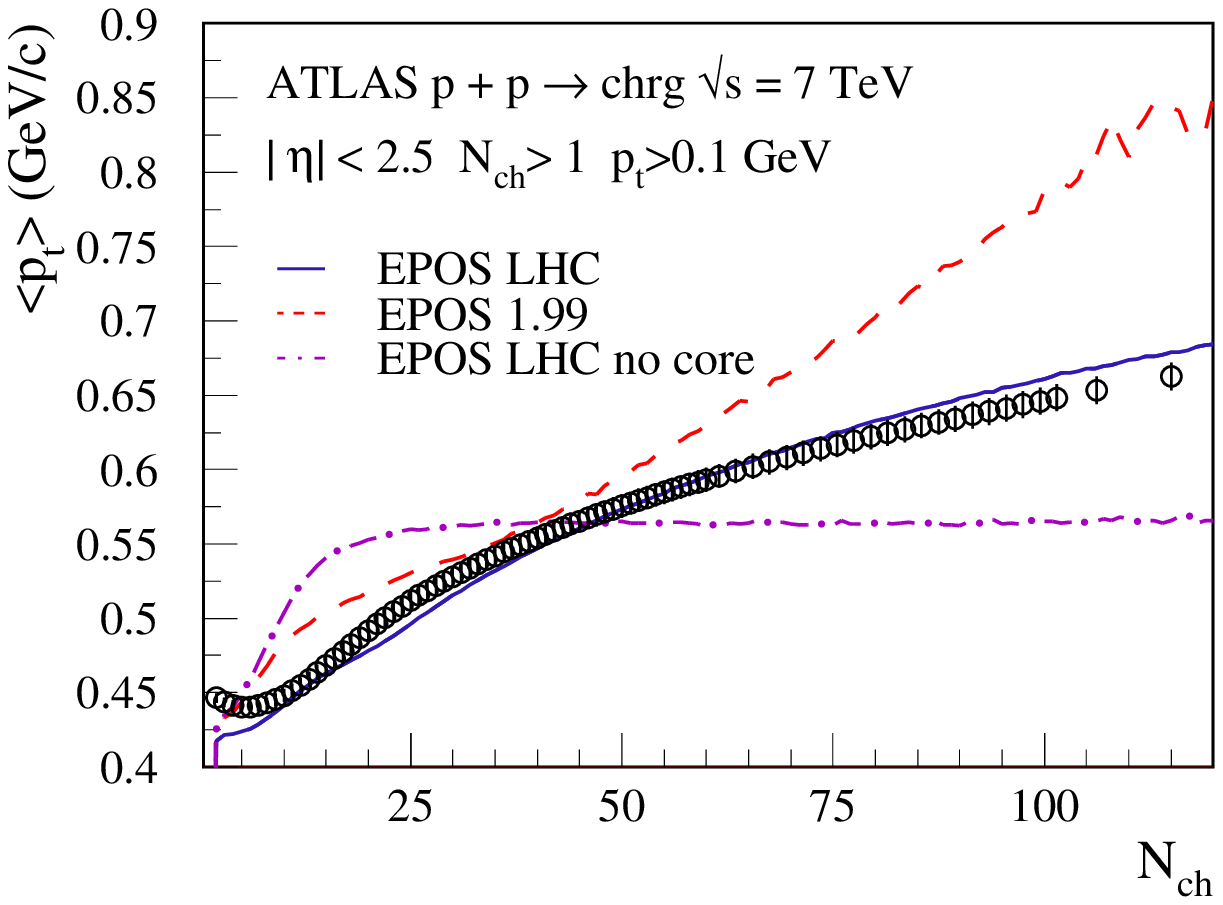}%
\caption{\label{apt7000} Average transverse momentum $\langle p_{\rm t}\rangle$
as a function of the number of charged particles for charged particles with
$p_{\rm t}>$0.1~GeV and $|\eta|<2.5$ for {\it p-p} interactions at 7~TeV. Simulations are done with EPOS~LHC (solid line), EPOS~1.99 (dashed line) and EPOS~LHC without core production (dash-dotted line). Points are data from ATLAS experiment~\protect\cite{Aad:2010ac}.}
\end{centering}
\end{minipage}
\hfill
\begin{minipage}[t]{0.48\linewidth}
\begin{centering}
\includegraphics[height=1.\columnwidth,angle=-90]{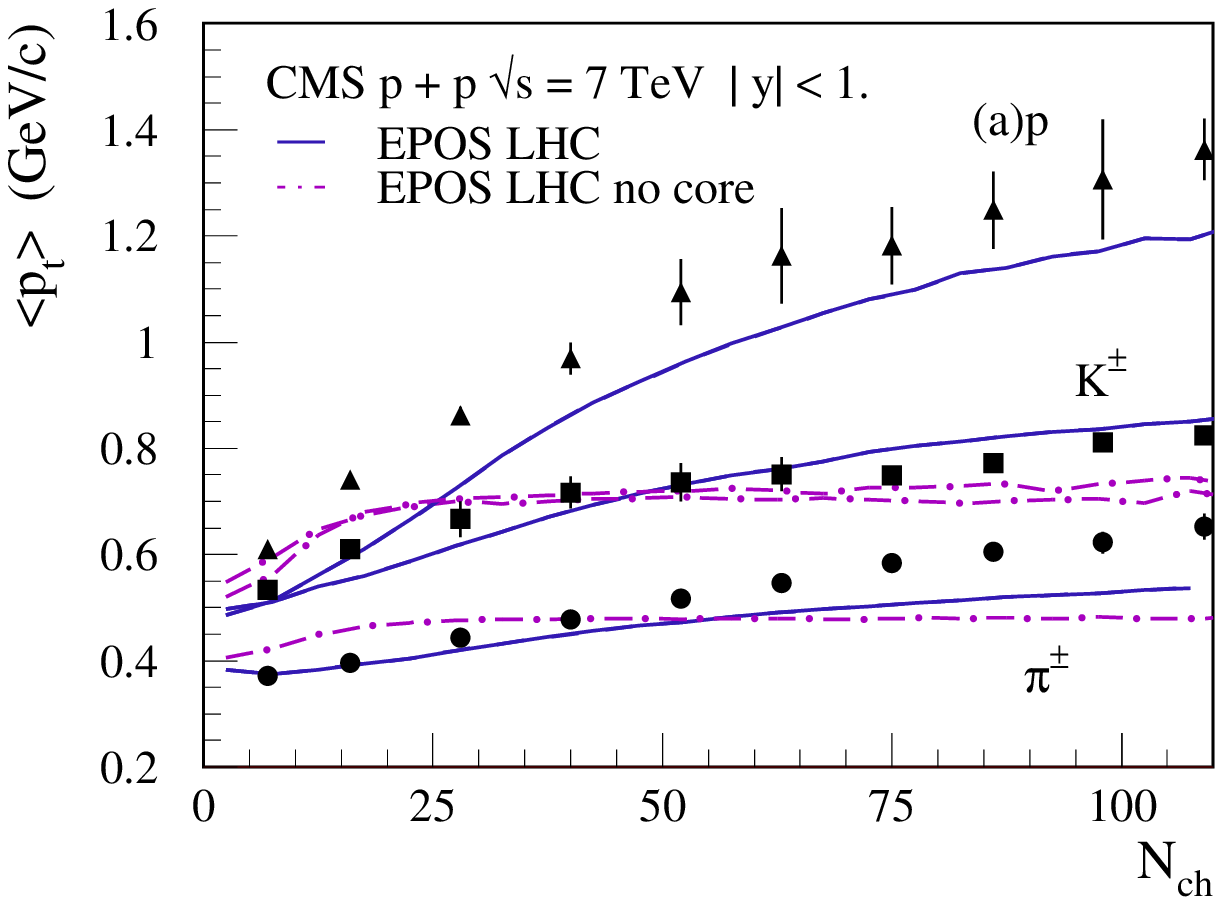}%
\caption{\label{aptid} Average transverse momentum $\langle p_{\rm t}\rangle$
of identified particles ($\pi$, {\it K} and {\it p}) as a function of the number of charged particles 
for particles with rapidity $|y|<1$ in {\it p-p} collisions at 7~TeV. Simulations 
are done with EPOS~LHC with (solid line) or without core (dash-dotted line).
Points are data from CMS experiment~\protect\cite{Chatrchyan:2012qb}.}
\end{centering}
\end{minipage}
\end{figure}

\begin{figure}[hptb]
\begin{centering}
\begin{minipage}[t]{0.7\linewidth}
\includegraphics[height=1.\columnwidth,angle=-90,bb=50 30 263 390]{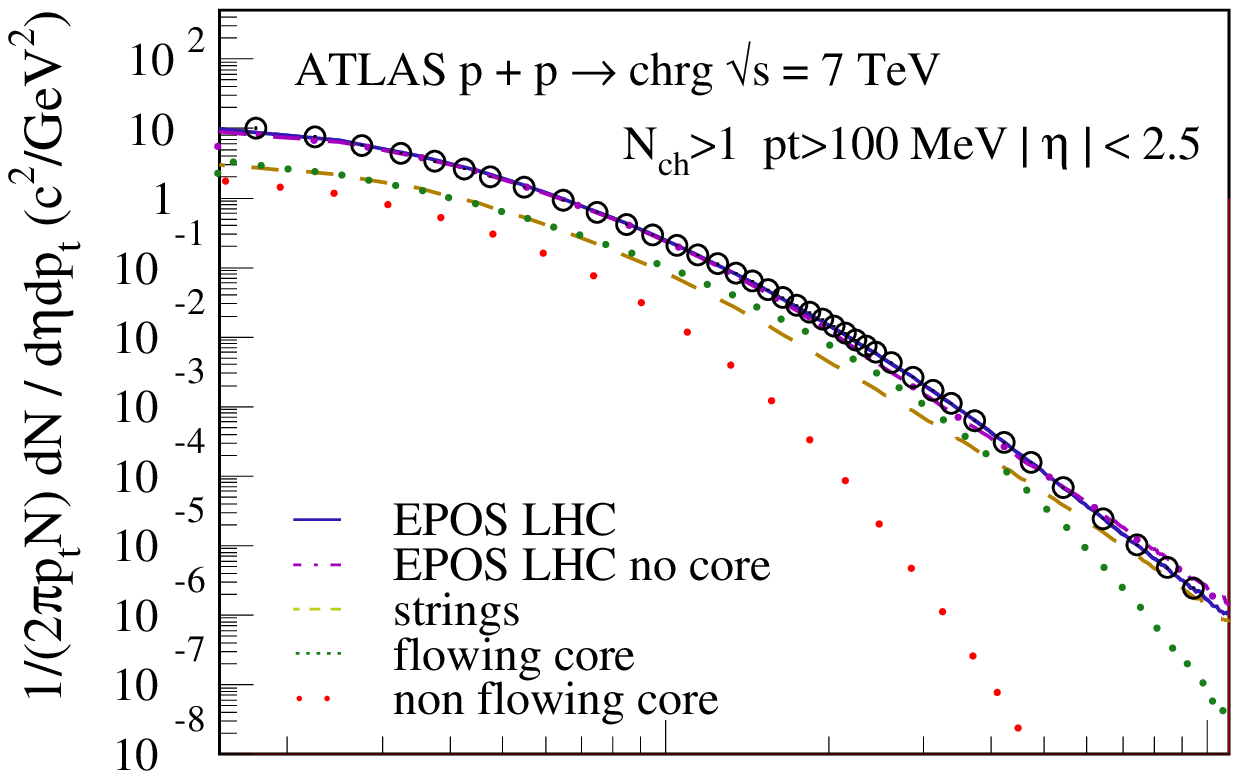}%
\hspace{-1.\columnwidth}
\includegraphics[height=1.\columnwidth,angle=-90]{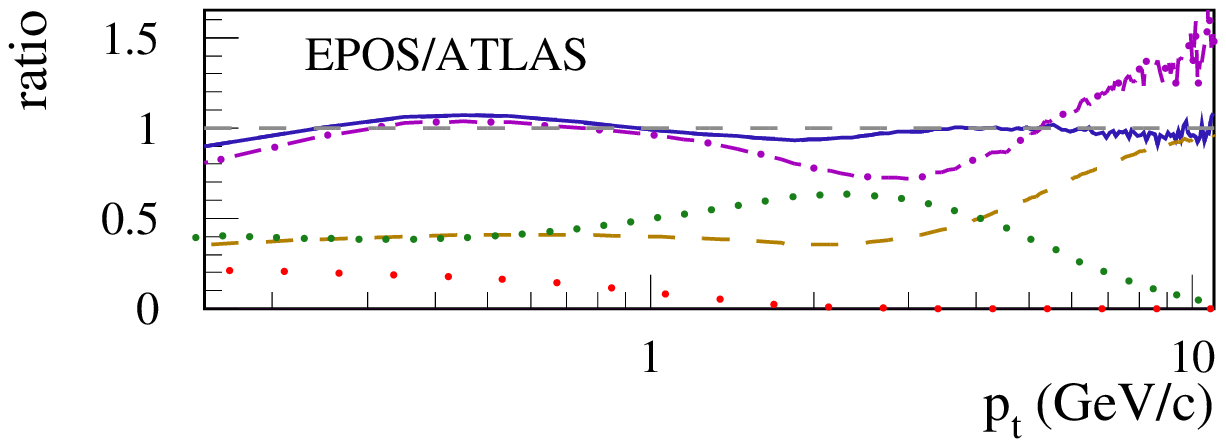}%
\end{minipage}
\caption{\label{pt}Transverse momentum $p_{\rm t}$ distribution of charged particles
with $|\eta|<2.5$ produced in minimum bias {\it p-p} collisions at 7~TeV. Simulations are done with EPOS~LHC with (solid line) or without core (dash-dotted line). 
The contributions of particles coming from the core hadronization are shown 
as dotted line while particles coming directly from string fragmentation are represented by a dashed line. Points are data from ATLAS experiment~\protect\cite{Aad:2010ac}.}
\end{centering}
\end{figure}

On figure~\ref{apt7000}, we first show that EPOS simulations without core 
formation (dash-dotted line) exhibit a flat behavior in case of hard
non-diffractive events ($N_{\rm ch}>25$). It is easy to understand in term of
string fragmentation even with multiple scattering since each string uses the 
same $\langle p_{\rm t}\rangle$. Then we can check that in the case of
EPOS~1.99 (dashed line) the $\langle p_{\rm t}\rangle$ due to radial flow was
extrapolated to too large value at 7~TeV. 
Using eq.~\ref{flowpp} in EPOS~LHC (solid line) and adjusting the parameter 
$y^{\rm px}$ to get the best fit at 900~GeV and 7~TeV of ATLAS data from 
\cite{Aad:2010ac}, it is possible to get a very good description of the 
measurements. Since the radial boost is based on a Lorentz transformation, it depends on
the total energy, thus on the mass, of each particle. The higher is the mass
the stronger will be the effect. It can be checked on the evolution
of the mean $p_{\rm t}$ as a function of the number of particles for identified
particles ($\pi$, {\it K} and {\it p}) as published by CMS experiment in~\cite{Chatrchyan:2012qb}. 
In figure~\ref{aptid} we can see that $\langle p_{\rm t}\rangle$ depends on the
mass of the particles and that EPOS~LHC give a reasonable description of the
data when the flow is active (solid lines) while the standard string
fragmentation gives a completely different behavior (dash-dotted line).

Comparing directly EPOS~LHC simulations with the transverse momentum
distribution measured by ATLAS experiment~\cite{Aad:2010ac} in minimum bias
{\it p-p} interaction like in figure~\ref{pt}, we can see that the particles coming
from the core hadronization with a radial flow (dotted line) will dominate 
the flux around $1-2$~GeV/c which is exactly the place where a deficit is 
observed when a model without flow (dash-dotted line) is compared to measured 
data. The position of the transition (and as a consequence of the shape of the
 $p_{\rm t}$ distribution) depends on the parameter $p_{\rm t}^{\rm cut}$ whose 
best value is 1~GeV/c. At large $p_{\rm t}$ ($>5$~GeV/c) the particles which 
are not completely absorbed into the high density region after string 
fragmentation dominate again.

\begin{figure}[hptb]
\begin{minipage}[t]{0.48\linewidth}
\begin{centering}
\includegraphics[height=1.\columnwidth,angle=-90]{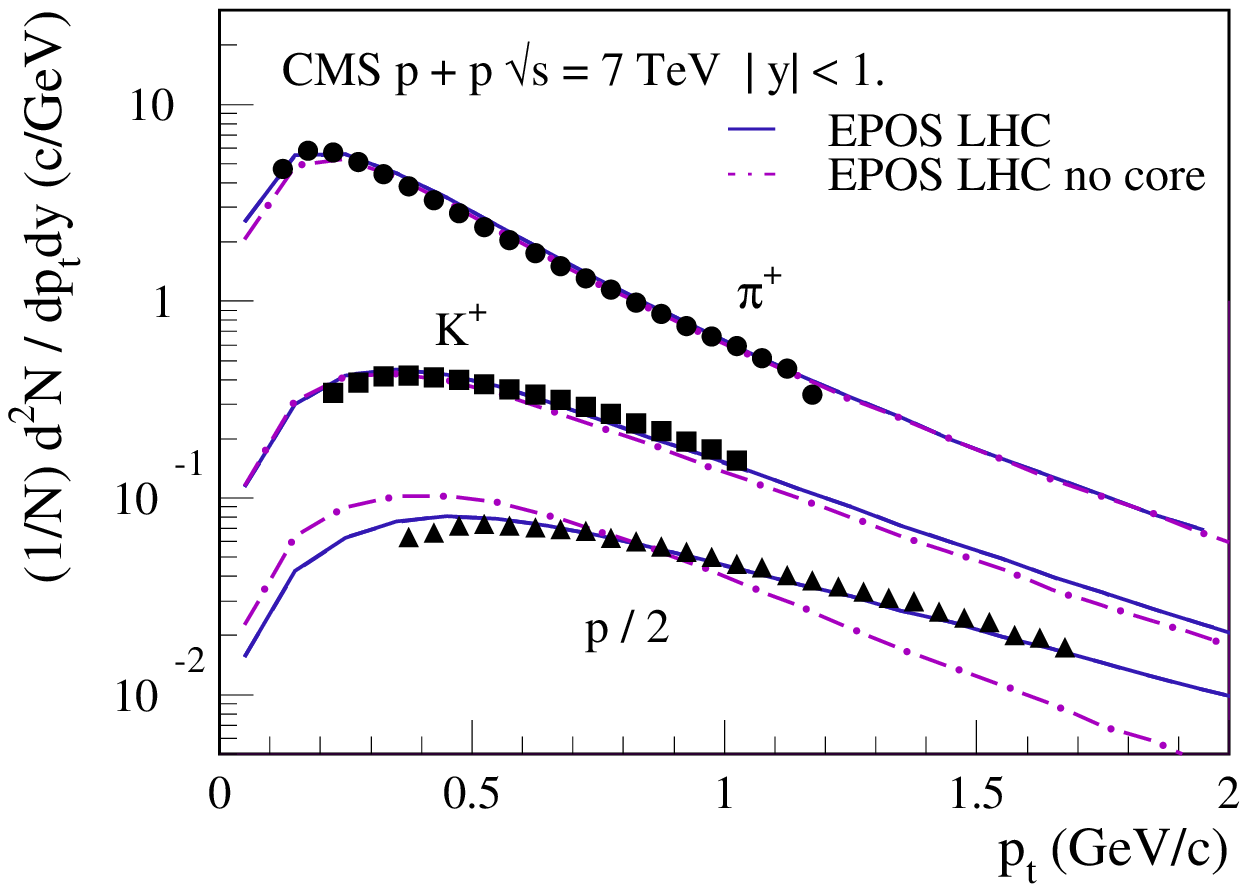}%
\caption{\label{ptid}Transverse momentum distribution of identified particles
($\pi$, {\it K} and {\it p} for $|y|<1$ for NSD {\it p-p} scattering at 7~TeV. Simulations are done with EPOS~LHC with (solid line) or without core (dash-dotted line). Points are data from CMS experiment~\protect\cite{Chatrchyan:2012qb}.}
\end{centering}
\end{minipage}
\hfill
\begin{minipage}[t]{0.48\linewidth}
\begin{centering}
\includegraphics[height=1.\columnwidth,angle=-90]{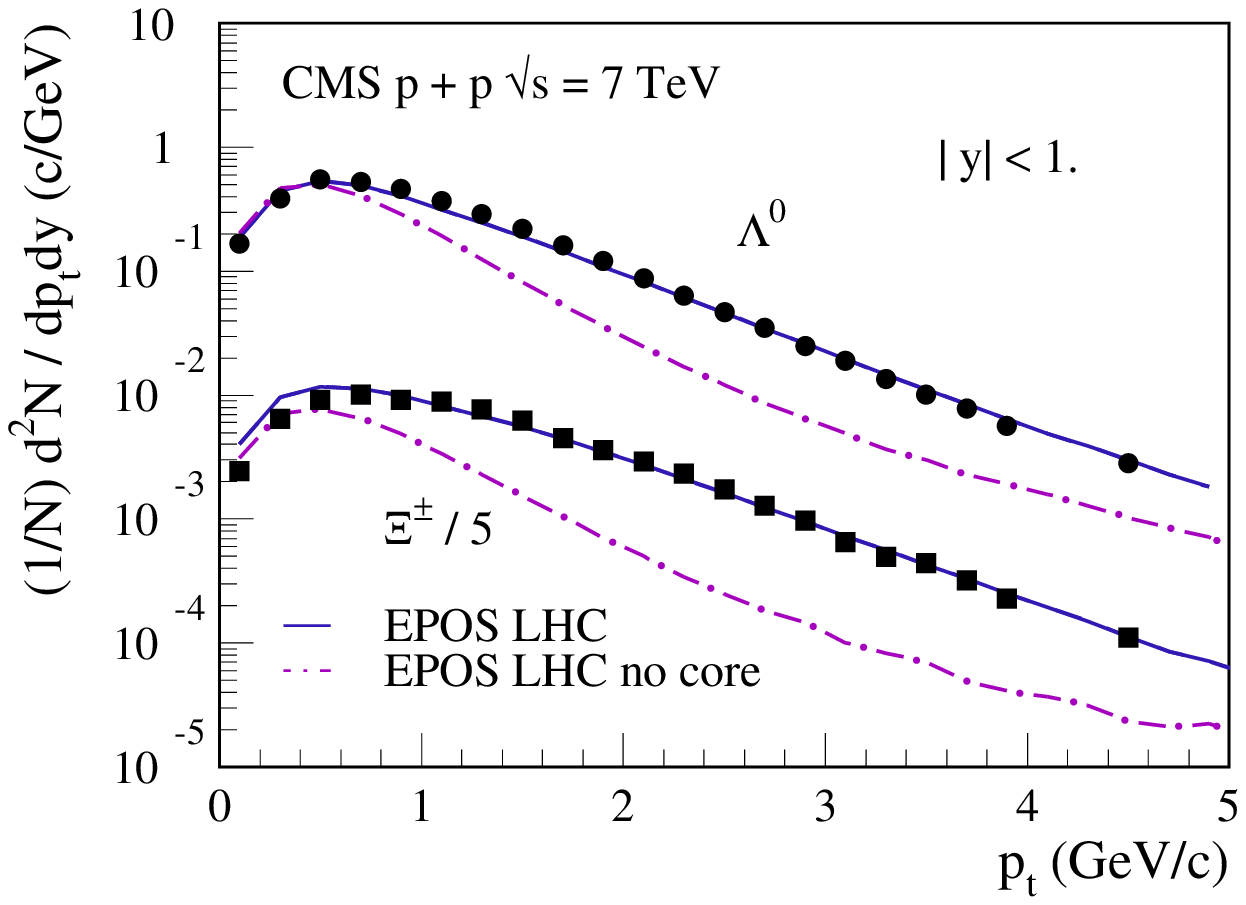}%
\caption{\label{ptstr}Transverse momentum distribution of strange baryons
($\lambda$ and $\xi$) for $|y|<1$ for NSD {\it p-p} scattering at 7~TeV. Simulations are done with EPOS~LHC with (solid line) or without core (dash-dotted line). Points are data from CMS experiment~\protect\cite{Khachatryan:2011tm}.}
\end{centering}
\end{minipage}
\end{figure}

In fact the effect is already clearly visible on minimum-bias transverse 
momentum distribution of identified particles. On figure~\ref{ptid} both
simulations with (solid line) or without (dash-dotted line) core formation
can describe $\pi$  $p_{\rm t}$ spectrum from~\cite{Chatrchyan:2012qb}. 
But when the mass increase, the 
deviation between the standard hadronization without flow and the data increase
while the simulations with collective hadronization give a good result. If
we consider strange baryons which has even larger masses, it can be seen on
figure~\ref{ptstr} that the difference between the two approaches can be as large
as a factor of 5 for Cascade particles~\cite{Khachatryan:2011tm} 
where the flow effect combines with the yield effect described in the 
previous section.

\begin{figure}[hptb]
\begin{centering}
\includegraphics[height=0.6\columnwidth,angle=-90]{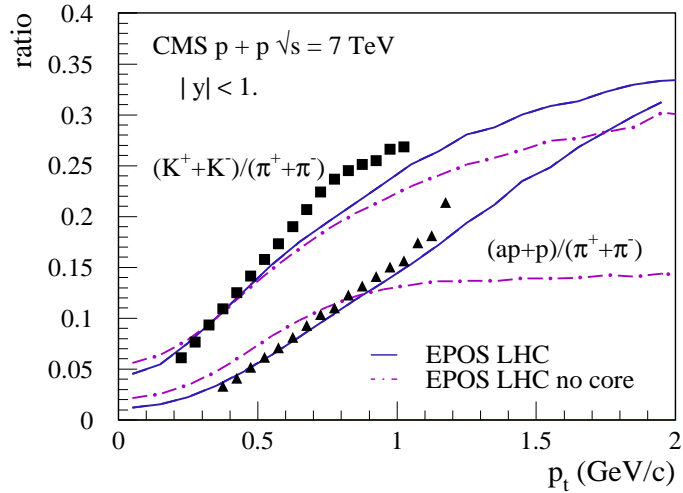}%
\caption{\label{rptid}Ratio of particle yield as a function of transverse momentum for $|y|<1$ for NSD {\it p-p} scattering at 7~TeV. Simulations are done with EPOS~LHC with (solid line) or without core (dash-dotted line). Points are data from CMS experiment~\protect\cite{Chatrchyan:2012qb}.}
\end{centering}
\end{figure}

Finally using the ratios of kaons over pions and proton over pions as a 
function of the transverse momentum as plotted on figure~\ref{rptid} it can
be clearly seen from data and simulations with (solid line) and without 
(dash-dotted line) that the flow effect take place only 
above $p_{\rm t}>1$~GeV/c.

\section{\label{ion}Heavy ion interactions}

The EPOS model was originally designed for heavy ion collisions. Even if 
the re-tuned LHC version described in this paper is based on the simplified
treatment of collective hadronization from~\cite{Werner:2007bf} and not on the
more sophisticated hydrodynamical treatment of~\cite{Werner:2010ny}, it is
important to check the basic distributions for heavy systems.

\subsection{Lead-Lead}

As explained in section \ref{collflow}, the so-called AA flow parametrization
reduces
cluster masses (and as a consequence the multiplicity of secondary particles)
 to increase the mean transverse momentum of the produced particles. As
a consequence to fix the parameter $y^{\rm mx}_{\rm rad}$ 
 both multiplicity and transverse momentum have to be taken
into account. 

\begin{figure}[hptb]
\begin{minipage}[t]{0.48\linewidth}
\begin{centering}
\includegraphics[height=1.\columnwidth,angle=-90]{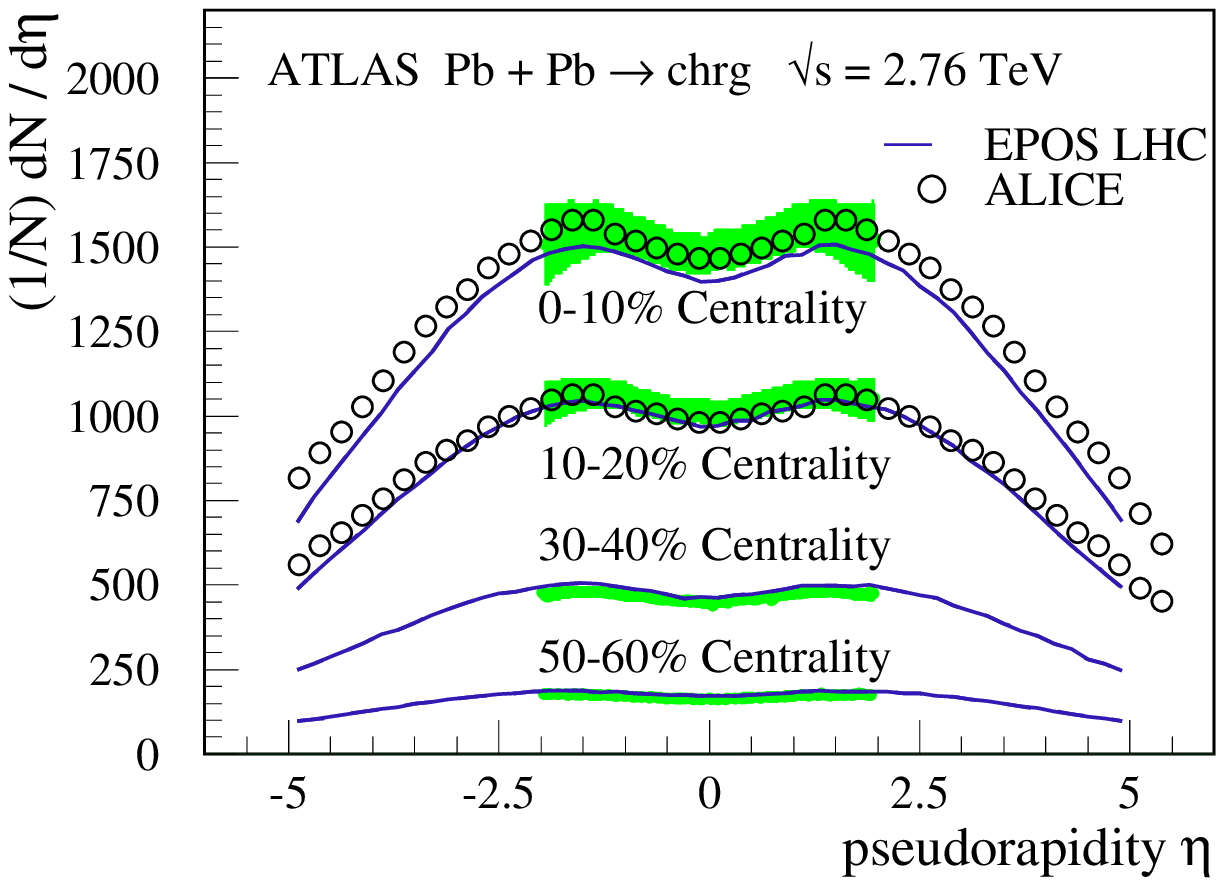}%
\caption{\label{etapbpb}Pseudorapidity distribution of charged particles in
{\it Pb-Pb} collisions at 2.76~TeV/A for centrality bins 0 to 10\%, 10 to 20\%,
30 to 40\% and 50-60\%. EPOS~LHC simulations (solid line) are compared to
ATLAS (band) and ALICE (point) measurements~\protect\cite{ATLAS:2011ag,Abbas:2013bpa}.}
\end{centering}
\end{minipage}
\hfill
\begin{minipage}[t]{0.48\linewidth}
\begin{centering}
\includegraphics[height=1.\columnwidth,angle=-90]{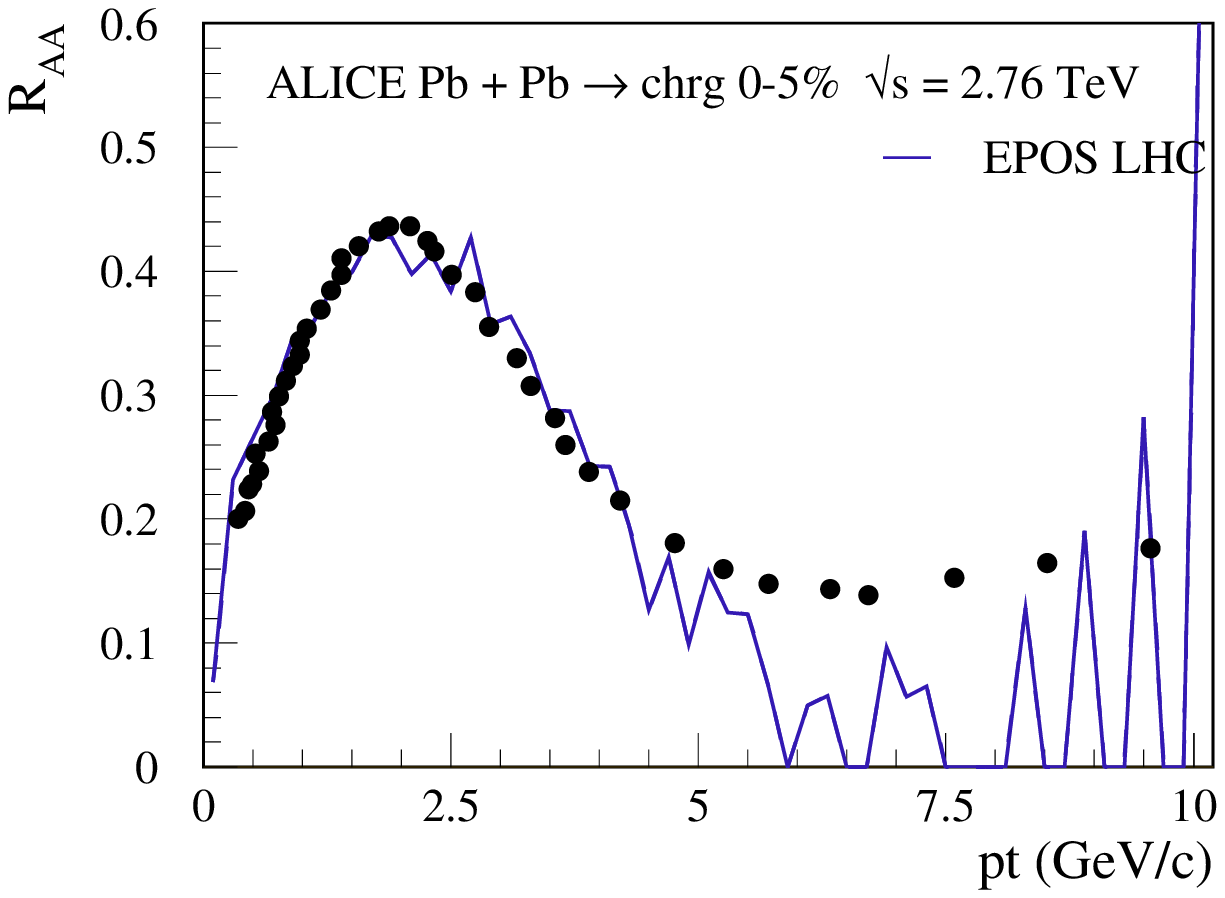}%
\caption{\label{rpbpb}Nuclear modification factor for charged particles
in most 5\% central events of {\it Pb-Pb} collisions at 2.76~TeV.EPOS~LHC 
simulations (solid line) are compared to ALICE 
measurements~\protect\cite{Aamodt:2010jd} (points).}
\end{centering}
\end{minipage}
\end{figure}

As it can be seen on figure~\ref{etapbpb} and~\ref{rpbpb}, it is possible
to achieve a good description
of both the pseudorapidity distribution of {\it Pb-Pb} collisions at 2.76~TeV and
 various centrality from the ATLAS and ALICE experiment~\cite{ATLAS:2011ag,Abbas:2013bpa} and of the
nuclear suppression factor of the most central events as measured by
 the ALICE experiment~\cite{Aamodt:2010jd} for $p_{\rm t}<5$~GeV/c. The large
suppression observed in the simulations with EPOS~LHC at larger $p_{\rm t}$ is
due to a lack of hard scattering during the initial stage of the
nucleus-nucleus interaction. Indeed the screening effects used in 
EPOS~\cite{Werner:2005jf} and necessary to have a good description of soft 
processes in {\it p-p} and {\it A-B} scattering affect hard scales the same way as
soft scales. In fact it has been shown now that such initial stage suppression
of hard processes is not observed in heavy ion data (gamma or Z boson 
production). This problem is being solved in the EPOS~3~\cite{Werner:2013tya} version
(currently under development).

\subsection{proton-Lead}

\begin{figure}[hptb]
\begin{minipage}[t]{0.48\linewidth}
\begin{centering}
\includegraphics[height=1.\columnwidth,angle=-90]{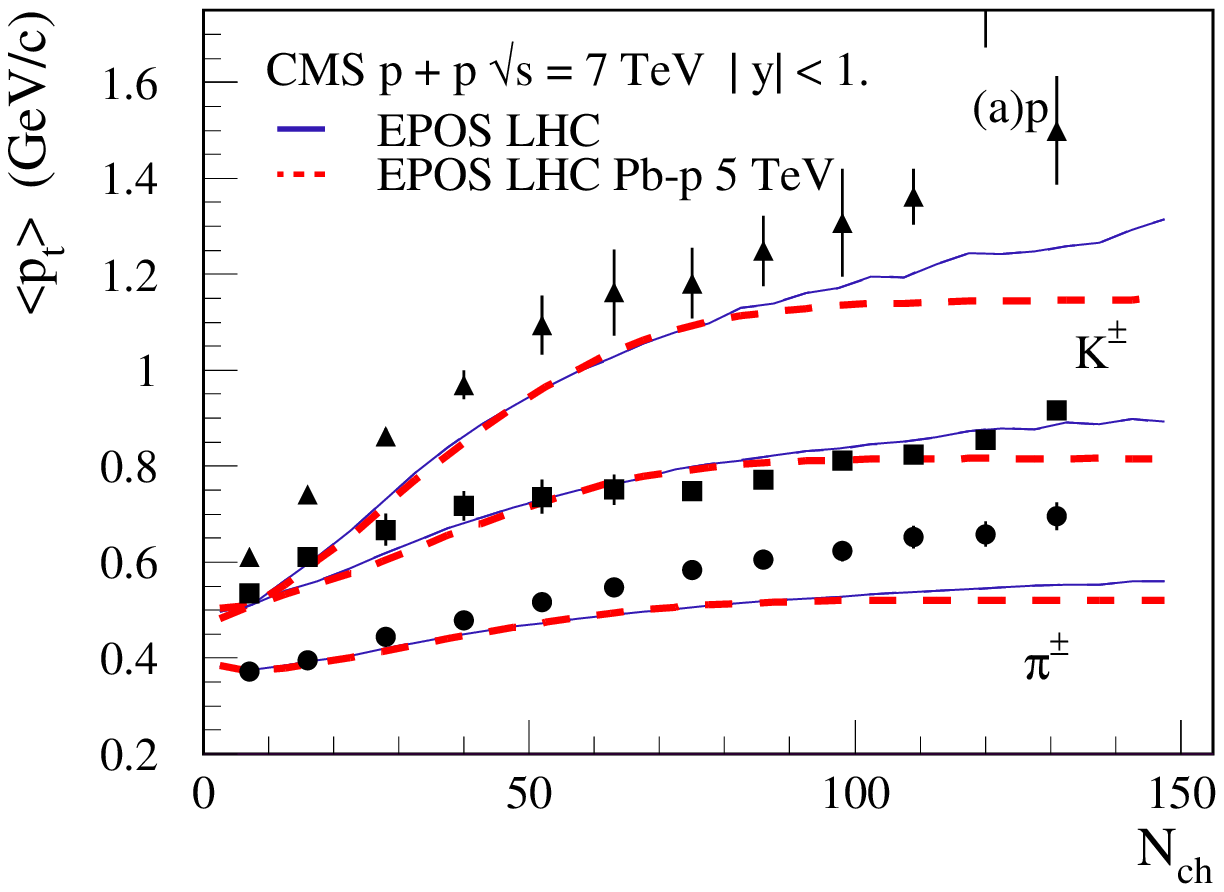}%
\caption{\label{aptidppb} Average transverse momentum $\langle p_{\rm t}\rangle$
of identified particles ($\pi$, {\it K} and {\it p}) as a function of the number of charged particles 
for particles with rapidity $|y|<1$ in {\it p-p} collisions at 7~TeV (solid line) 
and {\it Pb-p} collisions at 5~TeV (dashed line). Simulations 
are done with EPOS~LHC including core formation..
Points are data from CMS experiment~\protect\cite{Chatrchyan:2012qb} for
{\it p-p} scattering.}
\end{centering}
\end{minipage}
\hfill
\begin{minipage}[t]{0.48\linewidth}
\begin{centering}
\includegraphics[height=1.\columnwidth,angle=-90]{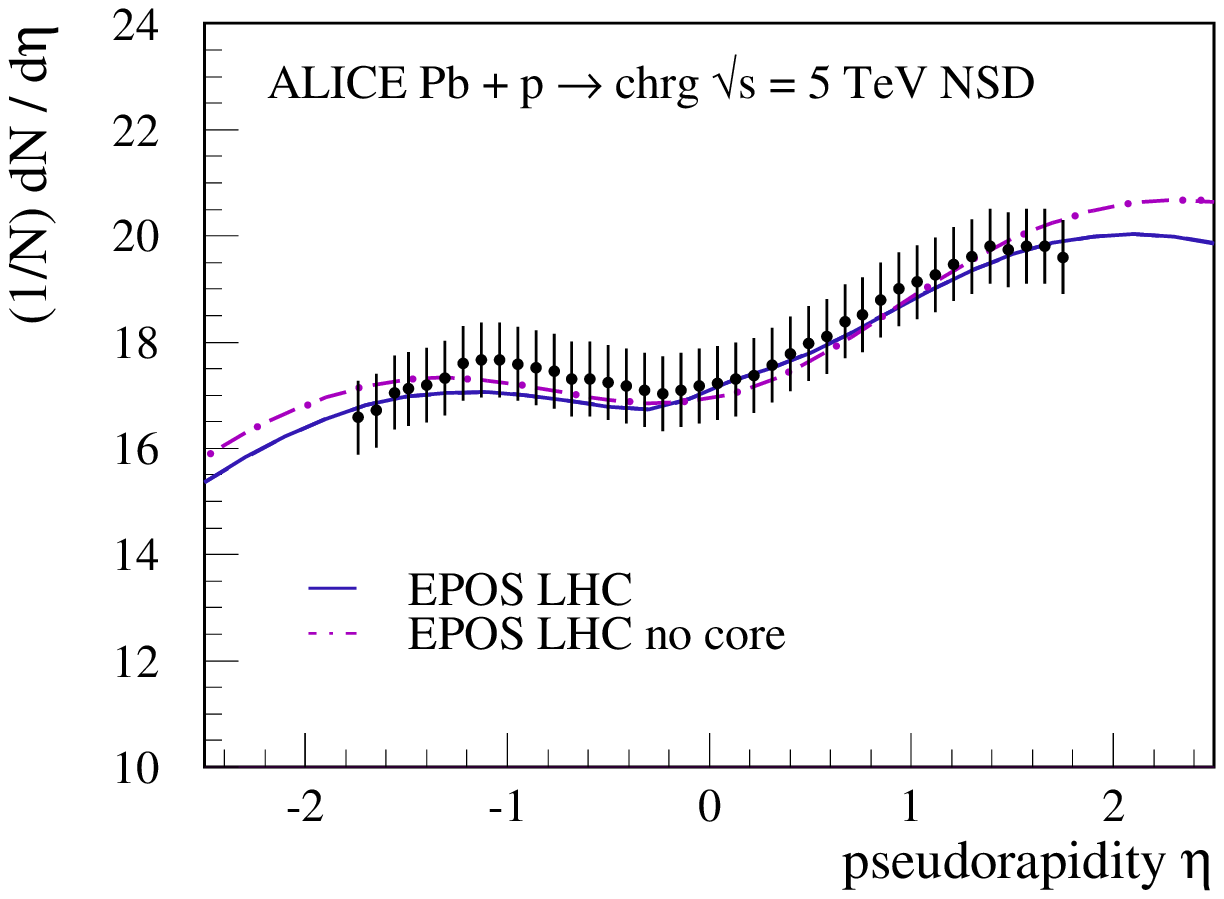}%
\caption{\label{etappb}Pseudorapidity distribution of 
charged particles from {\it Pb-p} collisions at 5.02~TeV. Simulations are done 
with EPOS~LHC with (solid line) or without core (dash-dotted line). 
Points are data from the ALICE experiment~\protect\cite{ALICE:2012xs}.}
\end{centering}
\end{minipage}
\end{figure}

From {\it p-p} and {\it p-Pb} data, all free parameters of eq.~\ref{eq:longflow}, 
\ref{eq:AAflow}, and \ref{eq:ppflow} are fixed. The free parameter in
eq.~\ref{eq:Mflow} is fixed in order to have the same flow in {\it p-p} and {\it p-A}
for the same multiplicity as shown on figure~\ref{aptidppb}. Here {\it Pb-p} 
simulations at 5~TeV (dashed line) are compared to data and simulations for 
{\it p-p} at 7~TeV like in figure~\ref{aptid}. At low multiplicity we observe the 
same flow behavior in {\it p-p} and {\it Pb-p} by construction (pp flow parametrization
 regime from eq.~\ref{flowpp}), but when N$_{ch}$ is higher than about
100 particles, the $\langle p_{\rm t}\rangle$ doesn't increase anymore because
we enter a different regime with a larger volume and we have a transition to
the AA flow parametrization (from eq.~\ref{eq:AAflow}). Since 
$y^{\rm px}_{\rm rad} \gg y^{\rm mx}_{\rm rad}$, the latter increase much slower
with the multiplicity.

To test the model predictions, it is now possible to compare to {\it Pb-p} data.
As we can see on figure~\ref{etappb}, the pseudorapidity distribution of 
charged particles from {\it Pb-p} collisions at 5.02~TeV as measured by the ALICE
experiment~\cite{ALICE:2012xs} is very well reproduced by EPOS~LHC (solid 
line). The effect of the core formation is very small on the average 
multiplicity (dash-dotted line without core). It is a real prediction since
no parameters has been changed to reproduce these data.

\begin{figure}[hptb]
\begin{minipage}[t]{0.48\linewidth}
\begin{centering}
\includegraphics[height=1.\columnwidth,angle=-90]{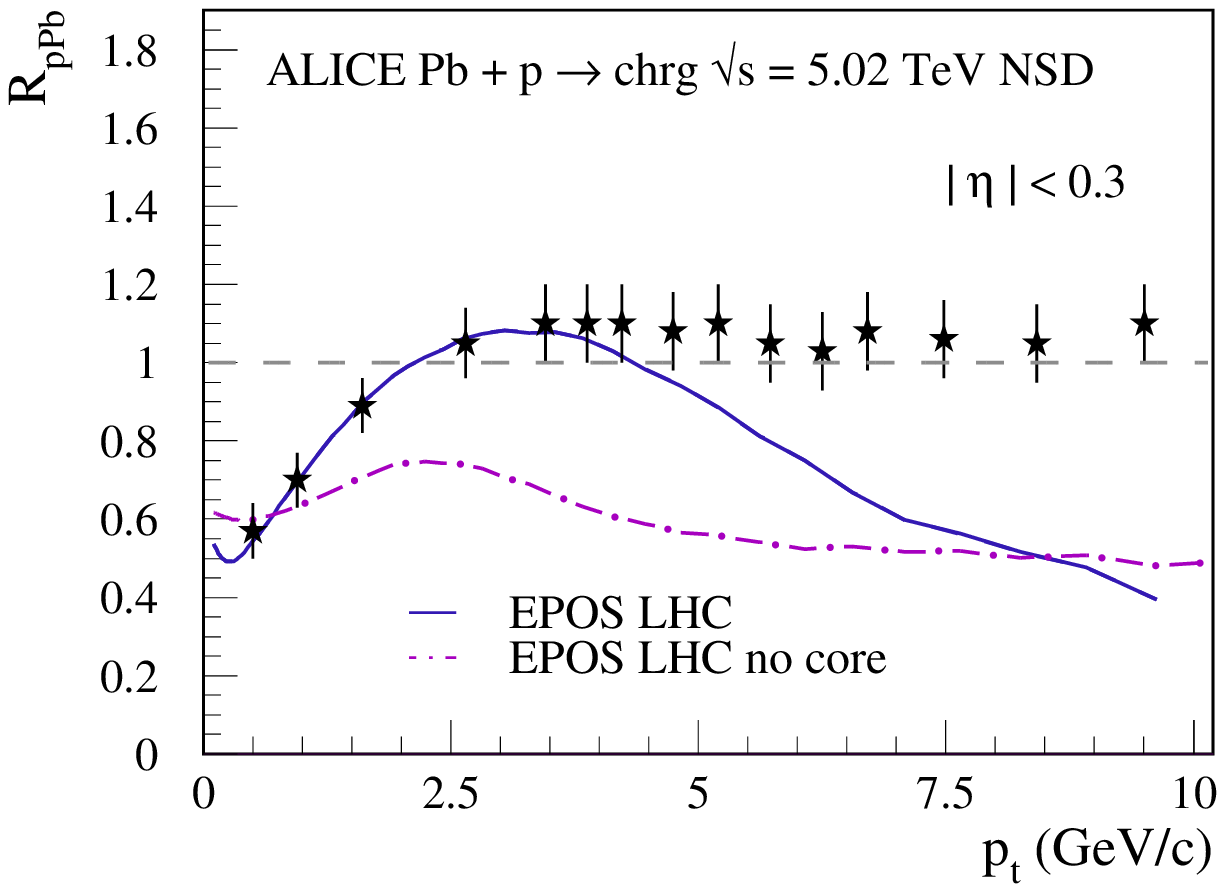}%
\caption{\label{rppb}Nuclear modification factor of 
charged particles from {\it Pb-p} collisions at 5.02~TeV. Simulations are done 
with EPOS~LHC with (solid line) or without core (dash-dotted line). 
Points are data from the ALICE experiment~\protect\cite{ALICE:2012mj}.}
\end{centering}
\end{minipage}
\hfill
\begin{minipage}[t]{0.48\linewidth}
\begin{centering}
\includegraphics[height=1.\columnwidth,angle=-90]{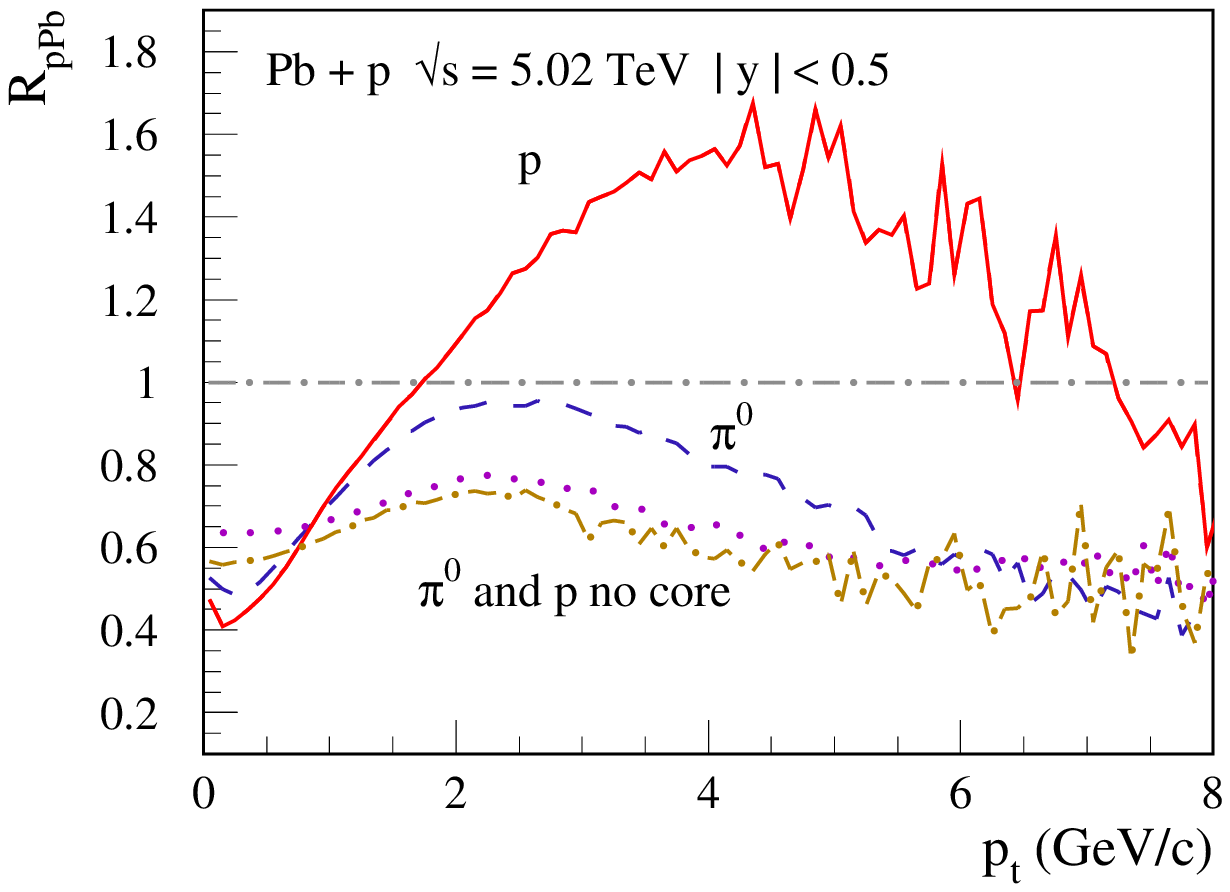}%
\caption{\label{rppbid}Nuclear modification factor for identified particles
($\pi^0$ (dashed and dash-dotted line) and proton {\it p} (solid and dashed line)
from {\it Pb-p} collisions at 5.02~TeV. Simulations are done 
with EPOS~LHC with (solid and dashed line) or without core (dotted and 
dash-dotted line).}
\end{centering}
\end{minipage}
\end{figure}

An important test of particle production in nuclear collisions is to study
the ratio of the $p_{\rm t}$ distribution in {\it p-A} with the one in {\it p-p} 
normalized by the number of binary collisions. It is called the nuclear
modification factor $R_{pPb}$. Any deviation from 1 indicates a nuclear effect. 
On figure~\ref{rppb} is presented the nuclear modification factor of 
charged particles from {\it Pb-p} collisions at 5.02~TeV measured by the ALICE
collaboration~\cite{ALICE:2012mj} together with EPOS~LHC simulations. Without
core formation (dash-dotted line) we have a constant $R_{pPb}=0.5$ due to 
the strong screening 
in nuclear collisions in EPOS which reduce the number of binary collision in
the initial state. This effect is important to get a correct multiplicity but
unfortunately the effect is the same for soft and hard processes leading to a
strong suppression of high $p_{\rm t}$ particles not observed in the data. If
the core formation is used (solid line), the situation improve a lot up to
$p_{\rm t}\sim 5$~GeV/c but then the strong suppression appears again (as it 
should be since the flow can not affect high $p_{\rm t}$ particles).

From the multiplicity measurement it is clear that there is a relative suppression of low $p_{\rm t}$ in {\it p-Pb}
relative to {\it p-p} and it is now clear that there is no suppression for
$p_{\rm t}>5$~GeV/c. But the transition region is dominated by the flow effect 
and the $R_{pPb}=1$ observed for $p_{\rm t}>2.5$~GeV/c has to be interpreted with
care because this value is probably unity by chance. If we compare the $R_{pPb}$
for different kind of particles (light $\pi^0$ and heavy proton {\it p}), we
can see on figure~\ref{rppbid} that EPOS predicts that both component will 
look completely different: mesons having $R_{pPb}<1$ and baryons $R_{pPb}>1$ due
to the stronger flow on heavy particles. This can be easily check on real data
and the effect will be even larger for multi-strange baryons.



\section{\label{pythia}Comparison with other minimum bias models}

In the following section EPOS~LHC is compared to the Pythia generator~\cite{Sjostrand:2006za}, which is commonly used to describe  hadron hadron collisions on an event-by-event basis for cms energies from Sp$\overline{p}$S to the LHC.  The FORTRAN based PYTHIA6 and the newer,  C++ based  Pythia8 version, implement very similar  soft QCD models, however the development of Pythia6 has stopped and new developments are only implemented in Pythia8.  
 Inclusive hadron-hadron collisions are modeled as a superposition of non-diffractive, single - and double diffractive processes, 
where the different 
 processes are mixed according to their cross section. The non-diffractive process is calculated at lowest order perturbative QCD two-to-two parton scatters, where the divergence at $p_t \rightarrow 0$ is regulated via a $p_{t,min}$ cut-off.
Multiple partonic scatters (MPI) are possible and their probability is calculated from the geometrical overlap function of the proton 
matter distribution.   Both the matter distribution and the $p_{t,min}$ cut-off are tuned to describe the minimum bias and underlying event data at different cms energies. The $p_t$ ordered parton shower algorithm is interleaved with the MPI. A different 
shower model using virtuality ordering for the emissions and  an older MPI model is also available in Pythia6, but was never 
tuned to LHC data. It is therefore not used for comparisons in this paper. 

Fragmentation is implemented via the Lund fragmentation model with free model parameters tuned to LEP data. The production 
of heavy quarks is suppressed according to $u : d : s  \sim 1: 1: 0.3 $, inspired by the quark masses. The production of $s \bar{s}$ production is a tunable parameter which is usually derived from LEP data. Charm and heavier quarks are not expected to be produced~\cite{Sjostrand:2006za}. 

The hadronization modeling includes a model of color reconnection \cite{Skands:2007zg}   which are implemented as re-arrangement of strings to minimize their length. Tuning of the free parameters in the color reconnection model yields a very good description of the relation between mean $p_t$ and the charged particle multiplicity as measured  at the LHC at a center-of-mass energy of 900 GeV and 7~TeV.

\begin{figure}[b]
\begin{minipage}[t]{0.48\linewidth}
\begin{centering}
\includegraphics[height=1.\columnwidth,angle=-90]{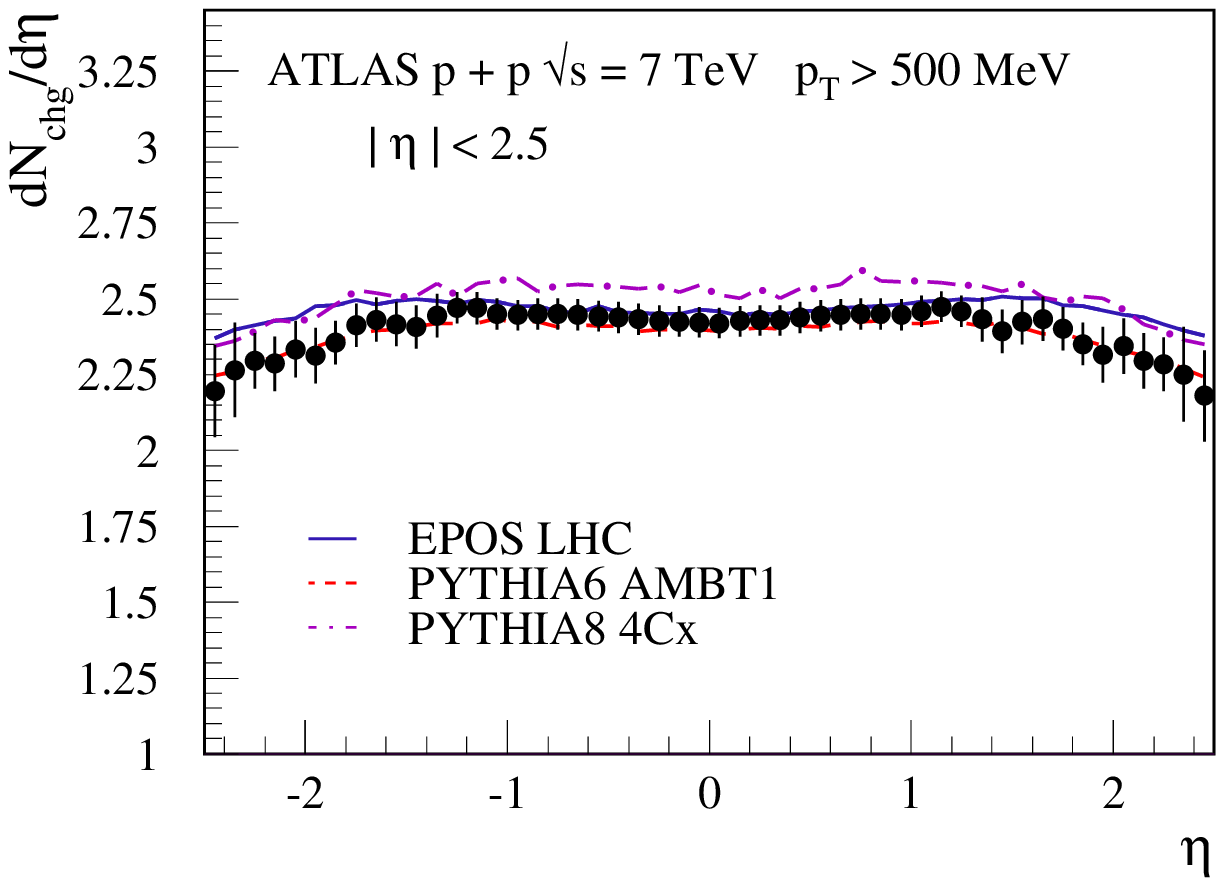}%
\caption{\label{fig_pythia_dndeta500} ATLAS measurement of the pseudorapidity distribution of charged particles with $p_t \geq 500$MeV in minimum bias  collisions~\protect\cite{Aad:2010ac} compared to  Pythia6 (dashed line), Pythia8 (dash-dotted line) and EPOS~LHC (solid line) simulations.}
\end{centering}
\end{minipage}
\hfill
\begin{minipage}[t]{0.48\linewidth}
\begin{centering}
\includegraphics[height=1.\columnwidth,angle=-90]{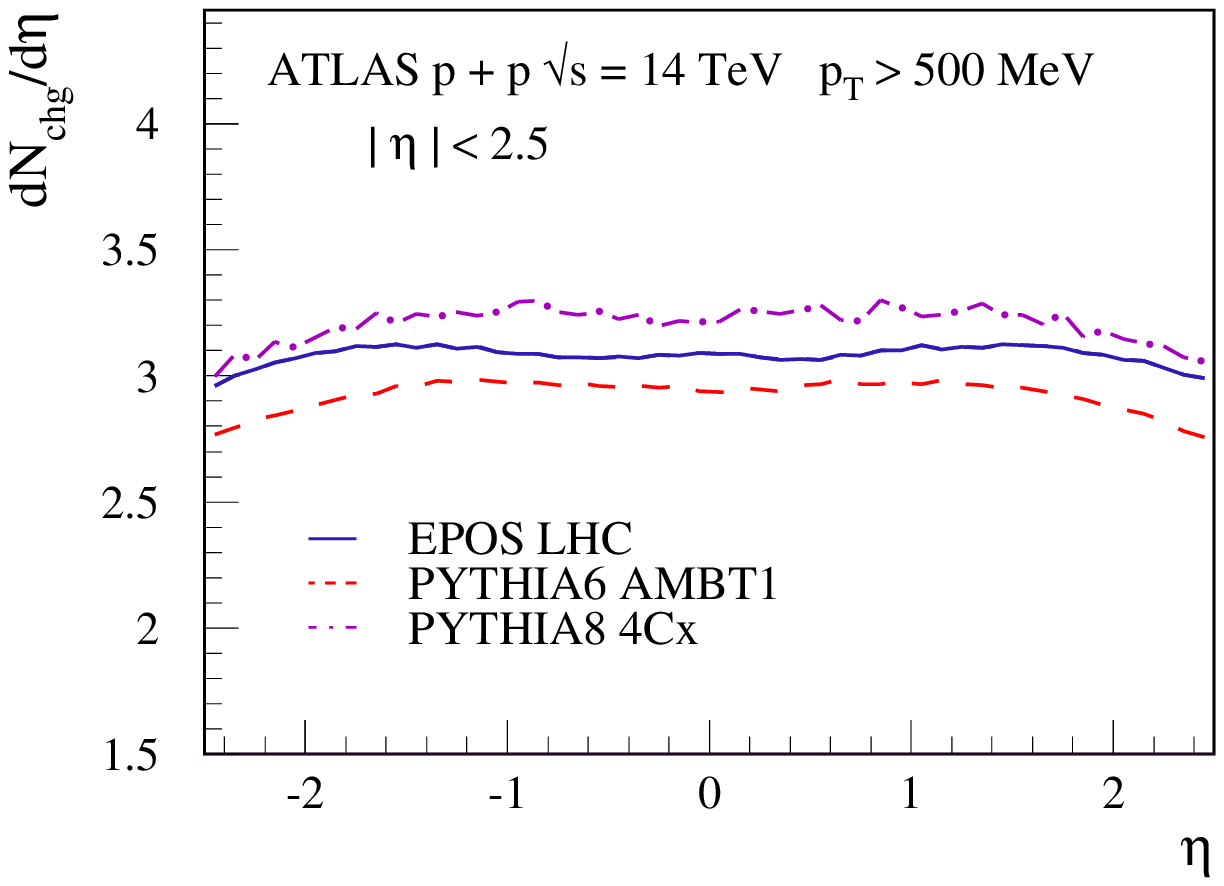}%
\caption{\label{fig_pythia_14_500}
 Predicted pseudorapidity distribution of charged particles $p_t \geq 500$MeV  in minimum bias  collisions at the LHC at a cms energy of 14~TeV compared from Pythia6 (dashed line), Pythia8 (dash-dotted line) and EPOS~LHC (solid line) simulations. }
\end{centering}
\end{minipage}
\end{figure}

 Soft diffraction is implemented within a Regge-based Pomeron model to generate the cross-section and the diffractive mass and momentum transfer \cite{Schuler:1993td,Schuler:1993wr} with some empirical corrections to cover the full phase space \cite{Sjostrand:2006za}. The soft diffraction model is extended in Pythia 8 with additional 
perturbative Pomeron-proton scattering using HERA diffractive PDFs \cite{Navin:2010kk}. 

Pythia  provides the possibility  to alter many modelling  details of the phenomenological soft QCD models. A large variety of tunes to minimum bias and underlying event data exist to optimize the descriptions. These tunes vary in the selection of model details, the parton density functions and  the observables they're tuned to, see  \cite{judithsreview} for a review of tunes and models.  Pythia in general yields a very good description of soft QCD effects both in minimum bias data and underlying event measurements at various center of mass energies. In the following we compare different LHC measurements to three different set-ups: to EPOS~LHC,  to Pythia6 with the AMBT1 tune \cite{ambt1} which was tuned to the ATLAS minimum bias  data and to  Pythia8 \cite{Sjostrand:2007gs} with  tune 4Cx \cite{4Cx} which was derived from comparison to   minimum bias and underlying event data from LHC. The plots were made using Rivet~\protect\cite{Buckley:2010ar} and are partially taken from MCPLOTS~\protect\cite{mcplots}.

Figure~\ref{fig_pythia_dndeta500} shows the charged particle production at LHC at  7~TeV with the requirement that the charged particles have a minimum $p_t$ of 500 MeV. The data are well described by the Pythia6 AMBT1 tune and  EPOS LHC which are both tuned to these data. It is also interesting to note, that the Pythia6 AMBT1 tune describes the full $\eta$ spectrum while the EPOS~LHC tune shows slight deviations in towards larger rapidity. The dependence of the charged particle production on the cms energy of the collision is  tunable in PYTHIA and this 
energy dependence has been derived from comparing to minimum bias data (and underlying event data for 4Cx) at 900~GeV and 7~TeV. Good agreement with the data can be reached for all models  also at 900~GeV\cite{mcplots}.  This also leads to  similar predictions for the yet unmeasured charged particle production at the LHC design energy of 14~TeV are  as shown in figure~\ref{fig_pythia_14_500}.

\begin{figure}[hptb]
\begin{centering}
\includegraphics[height=0.6\columnwidth,angle=-90]{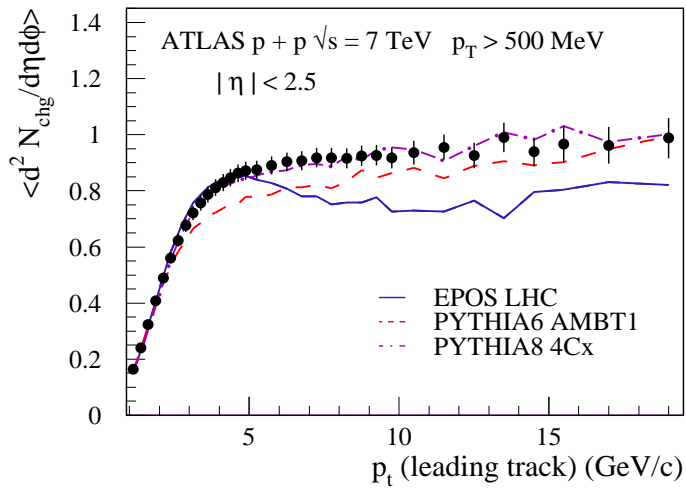}%
\caption{ \label{fig_pythia_ue500}
 ATLAS measurement  of  charged particles with $p_t \geq 500$MeV  produced in the underlying event, i.e. a region of 
$60^{\circ} \leq |\Delta \phi | \leq 120^{\circ} $ around the leading particle~\protect\cite{Aad:2010fh} compared to  Pythia6 (dashed line), Pythia8 (dashed-dotted line) and EPOS~LHC (solid line) simulations.  }
\end{centering}
\end{figure}

The slight deviation of Pythia8 4Cx in figure~\ref{fig_pythia_dndeta500} is caused by the fact that the model is tuned to  simultaneously  describe the 
underlying event data shown in figure~\ref{fig_pythia_ue500}. This  causes a slightly  too high prediction  of the particle production in minimum bias  events with the current Pythia models. A tune to the LHC  minimum bias  data set alone would give better description as demonstrated in \cite{ATLASpy8tuningNote}. Figure~\ref{fig_pythia_ue500} also shows that 
EPOS~LHC is not able to describe the soft particle production at the LHC when a hard scattering process  leading to 
a jet with $p_t$ above ~5 GeV is involved. This is again due to the fact that
that the non-linear effects as implemented in EPOS~LHC affect equally both soft and hard processes.

\begin{figure}[hptb]
\begin{minipage}[t]{0.48\linewidth}
\begin{centering}
\includegraphics[height=1.\columnwidth,angle=-90]{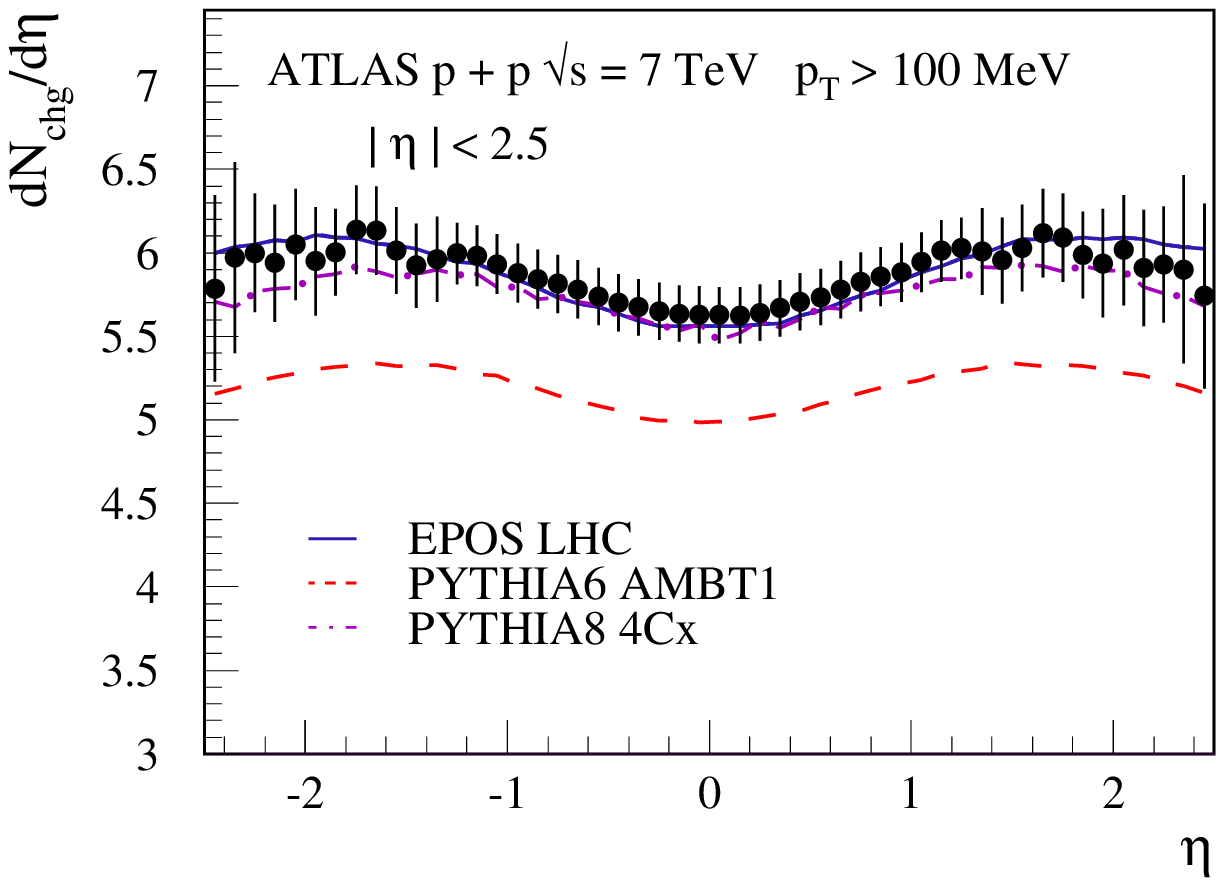}%
\caption{\label{fig_pythia_dndeta100}
 ATLAS measurement of the pseudorapidity distribution of  charged particles $p_t \geq 100$MeV  in minimum bias  collisions~\protect\cite{Aad:2010ac} compared to  Pythia6 (dashed line), Pythia8 (dash-dotted line) and EPOS~LHC (solid line) simulations.}
\end{centering}
\end{minipage}
\hfill
\begin{minipage}[t]{0.48\linewidth}
\begin{centering}
\includegraphics[height=1.\columnwidth,angle=-90]{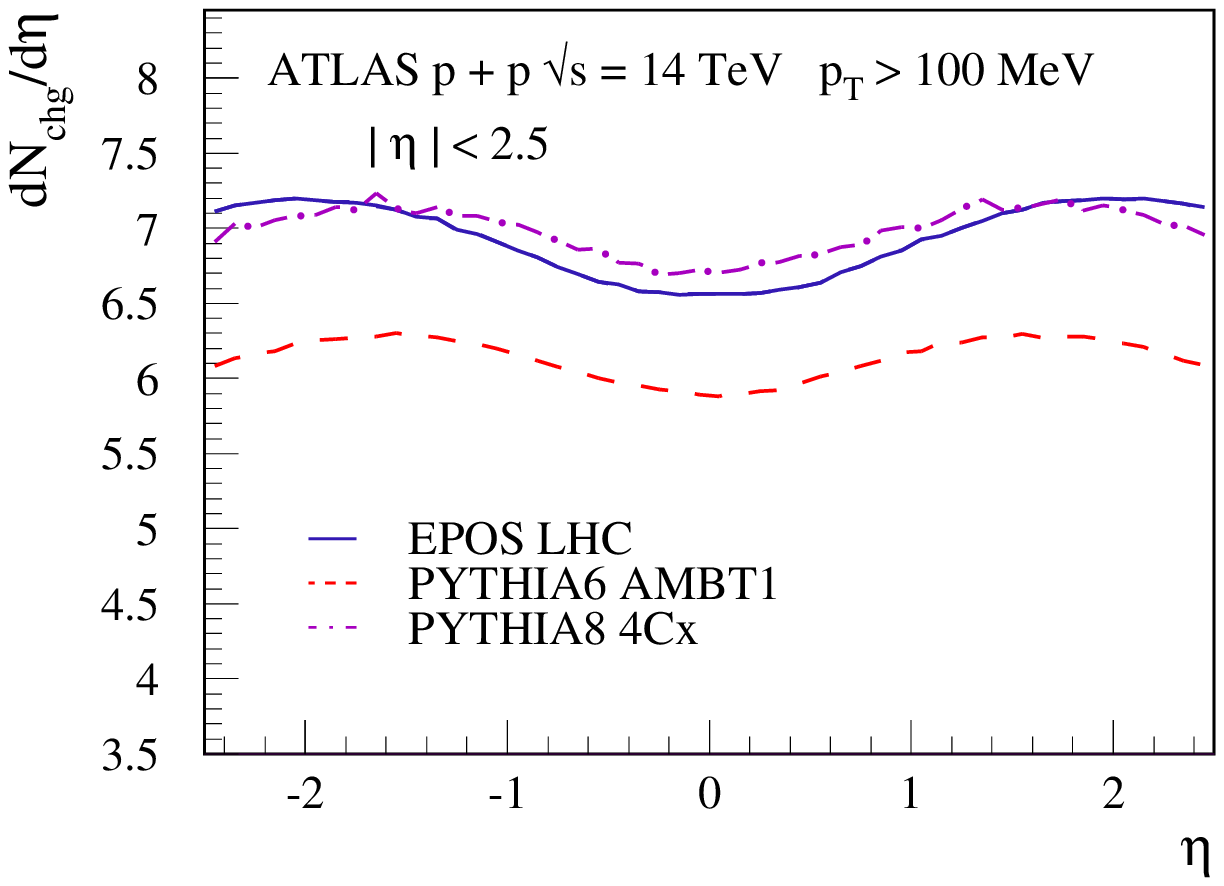}%
\caption{\label{fig_pythia_14_100}
 Predicted pseudorapidity distribution of charged particles $p_t \geq 100$MeV  in minimum bias  collisions at the LHC at a cms energy of 14~TeV compared from Pythia6 (dashed line), Pythia8 (dash-dotted line) and EPOS~LHC (solid line) simulations. }
\end{centering}
\end{minipage}
\end{figure}

Figure~\ref{fig_pythia_dndeta100} shows the same minimum bias data set, but  in this case includes charged particles with lower $p_t$, down to 100 MeV.  EPOS~LHC is able to describe both data sets with similar precision, however PYTHIA6 with AMBT1 shows significantly smaller increase in soft particle production than the data, leading to predictions which are significantly below the data. PYTHIA8 with tune 4Cx   describes the data well. It is worth to notice that the difference between PYTHIA6 and EPOS increases if we compare the same distribution at 14~TeV, see figure\ref{fig_pythia_14_100}. 

\begin{figure}[hptb]
\begin{minipage}[t]{0.48\linewidth}
\begin{centering}
\includegraphics[height=1.\columnwidth,angle=-90]{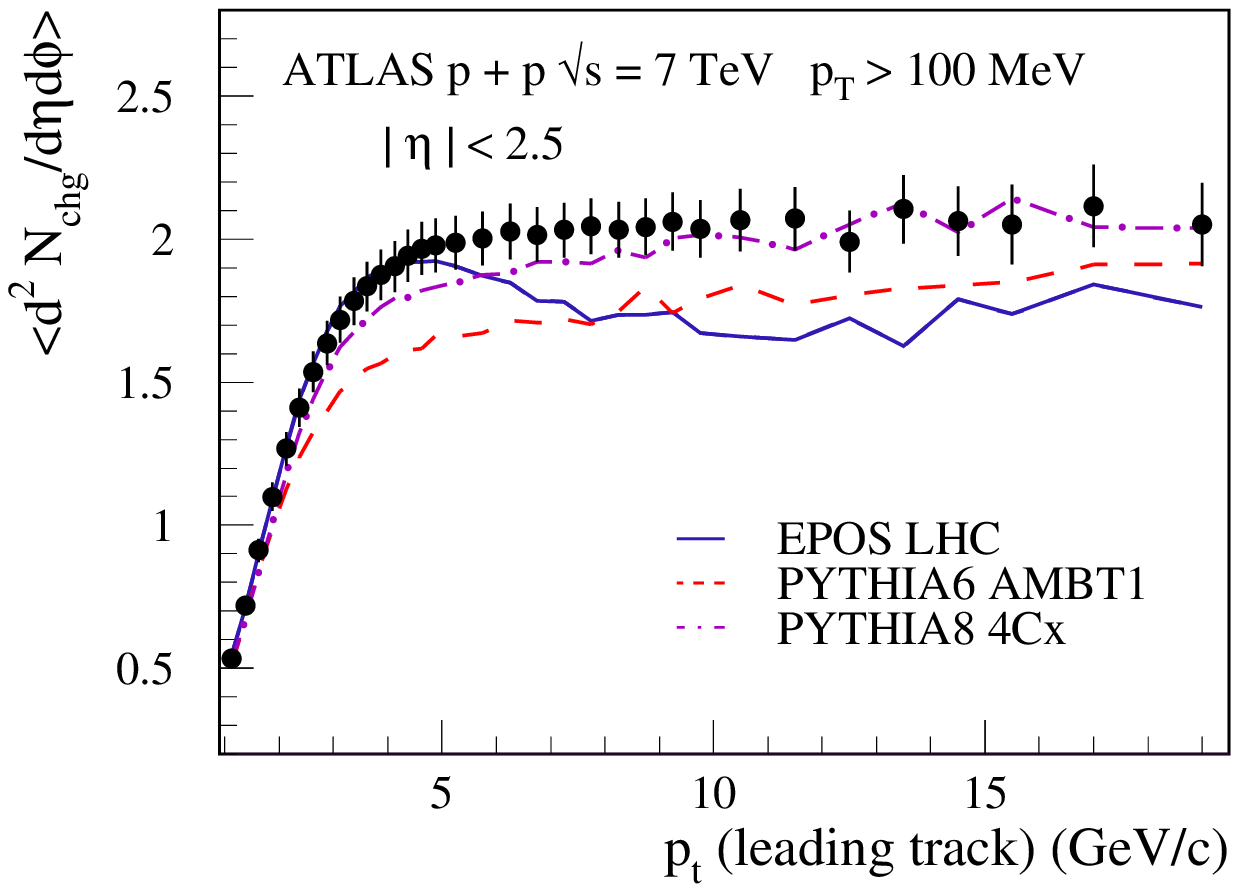}%
\caption{ \label{fig_pythia_ue100}
 ATLAS measurement  of  charged particles with $p_t \geq 100$MeV  produced in the underlying event, i.e. a region of 
$60^{\circ} \leq |\Delta \phi | \leq 120^{\circ} $ around the leading particle~\protect\cite{Aad:2010fh} compared to  Pythia6 (dashed line), Pythia8 (dash-dotted line) and EPOS~LHC (solid line) simulations.  }
\end{centering}
\end{minipage}
\hfill
\begin{minipage}[t]{0.48\linewidth}
\begin{centering}
\includegraphics[height=1.\columnwidth,angle=-90]{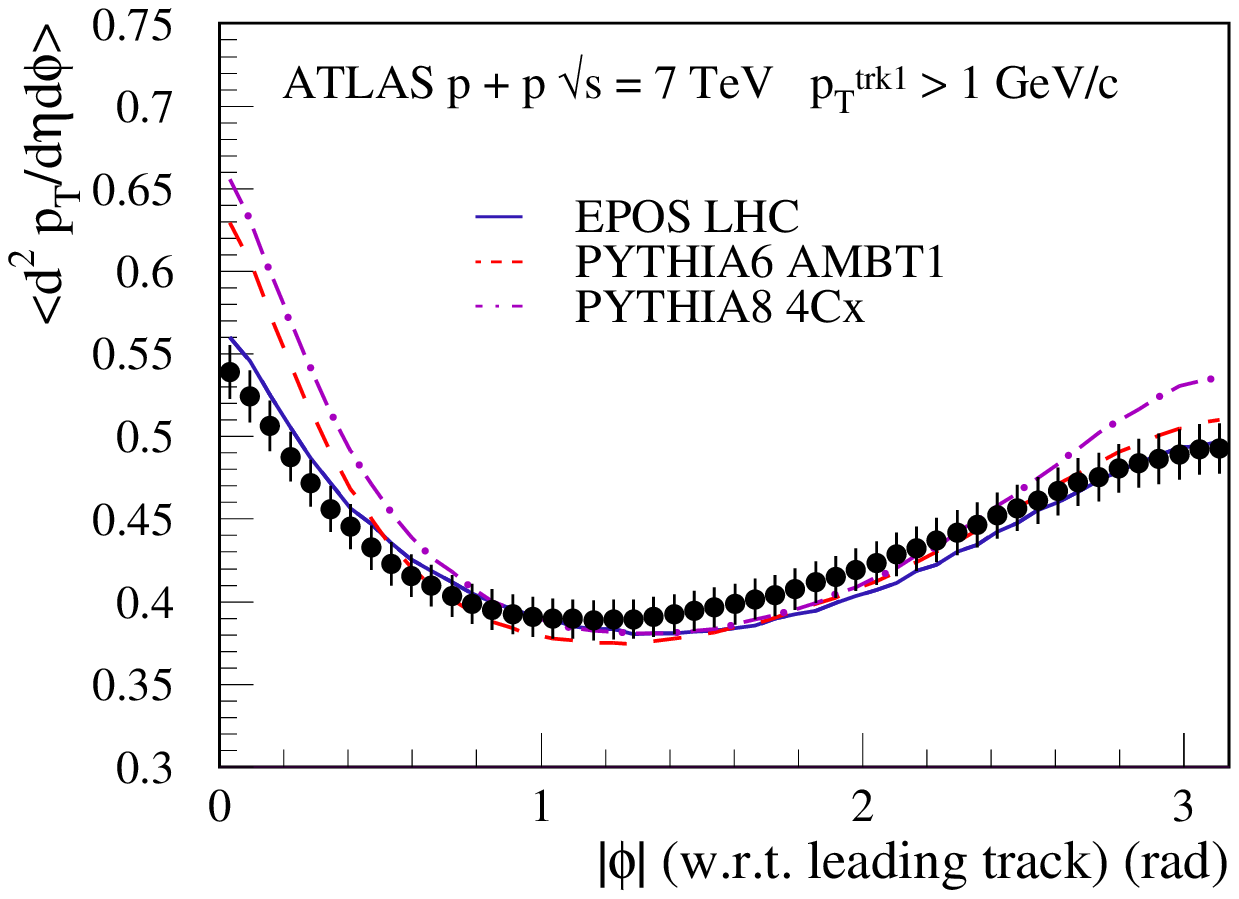}%
\caption{\label{fig_pythia_dphi} ATLAS measurement  of the angular distribution of charged particles production with respect to the leading charged particle with   $p_t \geq 1$GeV~\protect\cite{Aad:2010fh} compared to  Pythia6 (dashed line), Pythia8 (dash-dotted line) and EPOS~LHC (solid line) simulations.   }
\end{centering}
\end{minipage}
\end{figure}

The angular distribution of the charged particle production has also been measured, where the angle is measured with 
respect to the leading particle. This observable is sensitive to the emergence of jets or so-called mini-jets in hadronic collisions.
ATLAS has measured this distribution with various thresholds for  the leading particle $p_t$, ranging from 1. to 5 GeV.  figure~\ref{fig_pythia_dphi} shows the distribution for the lowest $p_t$ threshold, where  EPOS~LHC is the only generator 
to provide a very good description both of the total amount and of the shape of the distribution. As the leading particle $p_t$ increases, the description of the data by the pythia models improves, but has up to 20\% deviations close to the leading particle. 
The description of the data by EPOS~LHC gets slightly worse for the higher $p_t$ thresholds, but is always within 10\%.

\begin{figure}[hptb]
\begin{centering}
\begin{minipage}[t]{0.6\linewidth}
\includegraphics[height=1.\columnwidth,angle=-90,bb=50 30 271 390]{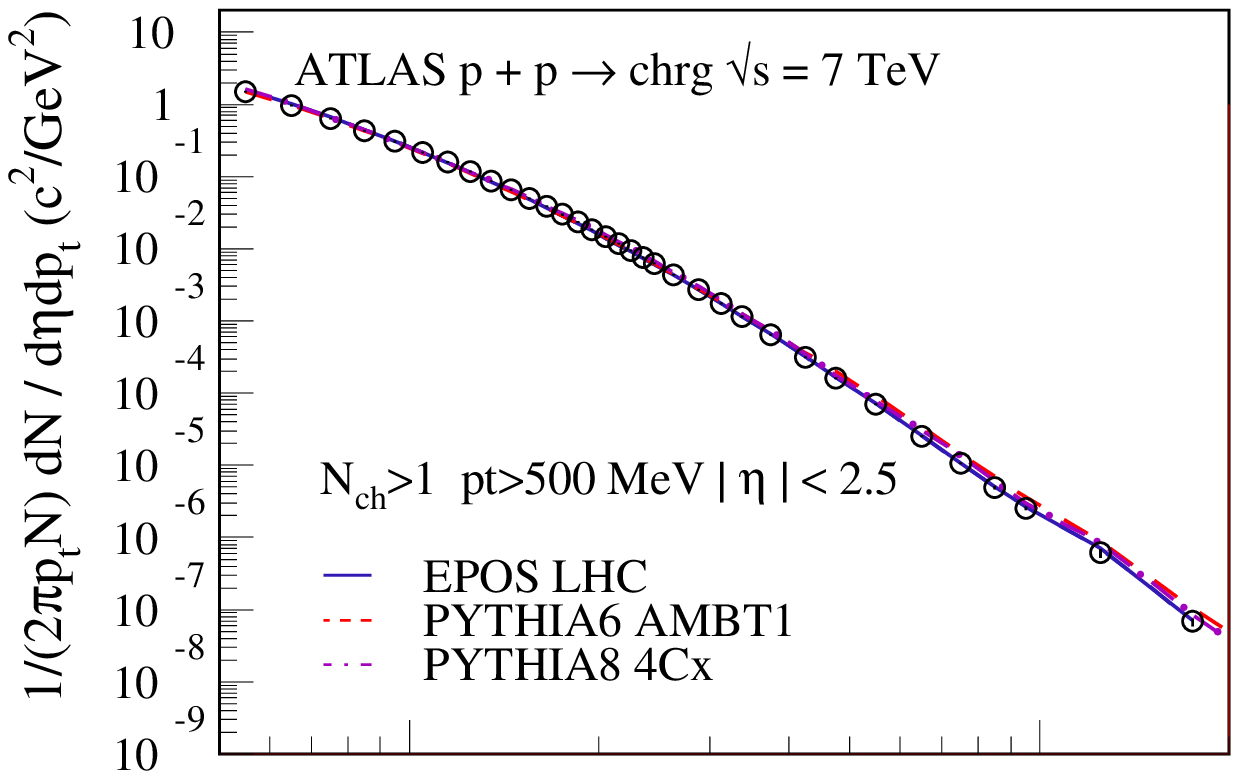}%
\hspace{-1.\columnwidth}
\includegraphics[height=1.\columnwidth,angle=-90]{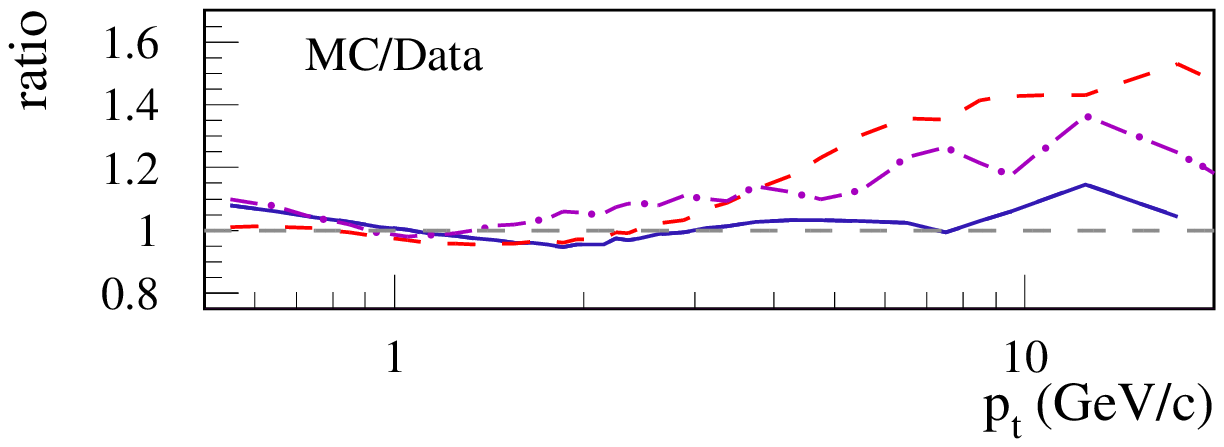}%
\end{minipage}
\caption{\label{fig_pythia_pt} ATLAS measurement of $p_t$ of charged particles produced in minimum bias collisions at 7~TeV~\protect\cite{Aad:2010ac} compared to  Pythia6 (dashed line), Pythia8 (dash-dotted line) and EPOS~LHC (solid line) simulations.  }
\end{centering}
\end{figure}

The $p_t$ spectrum of particles in minimum bias events is particularly sensitive to collective flow effects and  
has been measured by ATLAS and CMS~\cite{Aad:2010ac, Khachatryan:2010us}. As shown in figure~\ref{fig_pythia_pt}, 
both pythia variants have problems to describe the  $p_t$ spectrum above ~2GeV, despite the effort  to tune to these distributions, it seems that the Pythia tunes miss an aspect to get a high precision description of these spectra.  EPOS~LHC tune describes this specrum well due to the core with its collective flow, see figure~\ref{pt}.

\begin{figure}[hptb]
\begin{minipage}[t]{0.48\linewidth}
\begin{centering}
\includegraphics[height=1.\columnwidth,angle=-90]{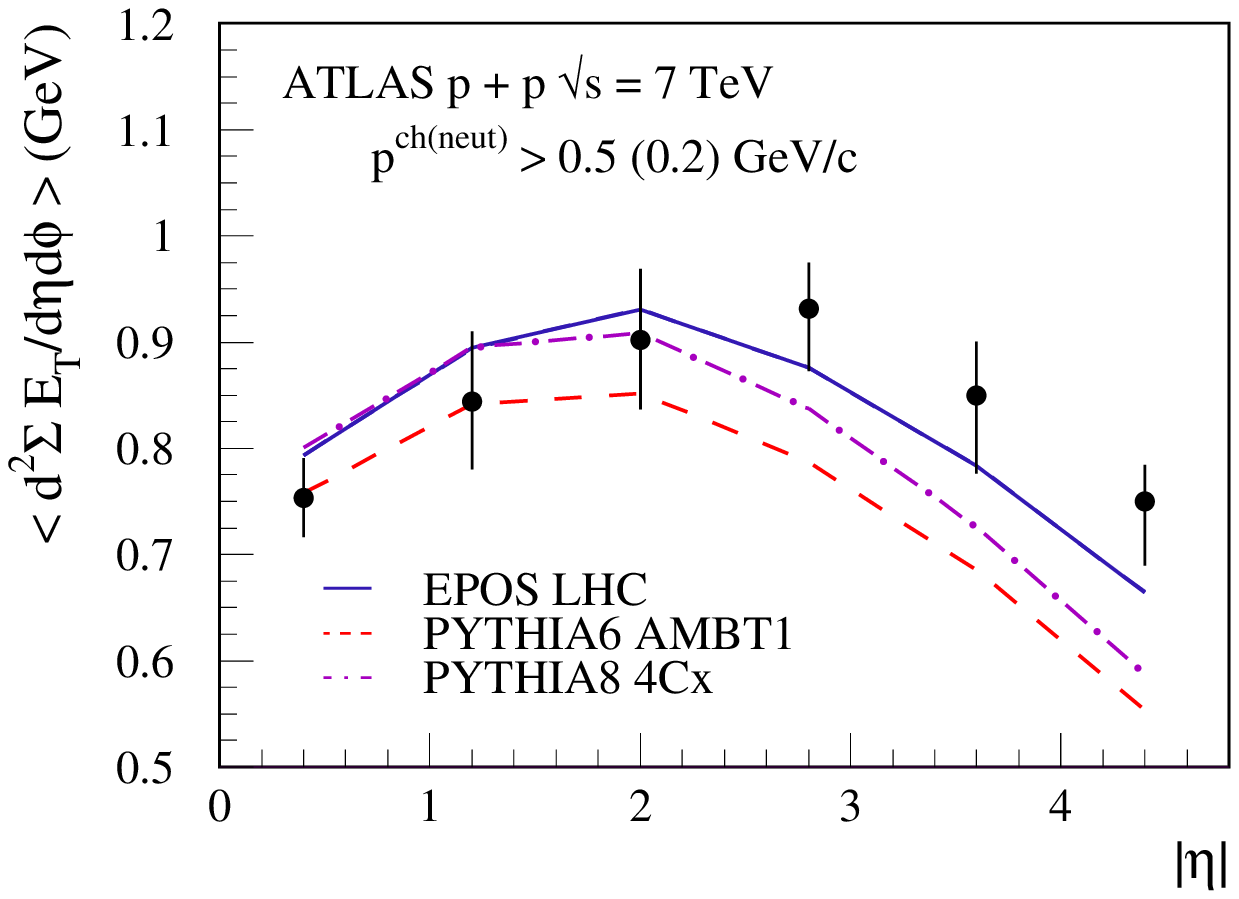}%
\caption{ \label{fig_pythia_sumEtvseta} ATLAS measurement of $\sum{E_t}$ as a function of pseudorapidity $\eta$ in minimum bias events at 7~TeV~\protect\cite{Aad:2012mfa} compared to  Pythia6 (dashed line), Pythia8 (dash-dotted line) and EPOS~LHC (solid line) simulations. }
\end{centering}
\end{minipage}
\hfill
\begin{minipage}[t]{0.48\linewidth}
\begin{centering}
\includegraphics[height=1.\columnwidth,angle=-90]{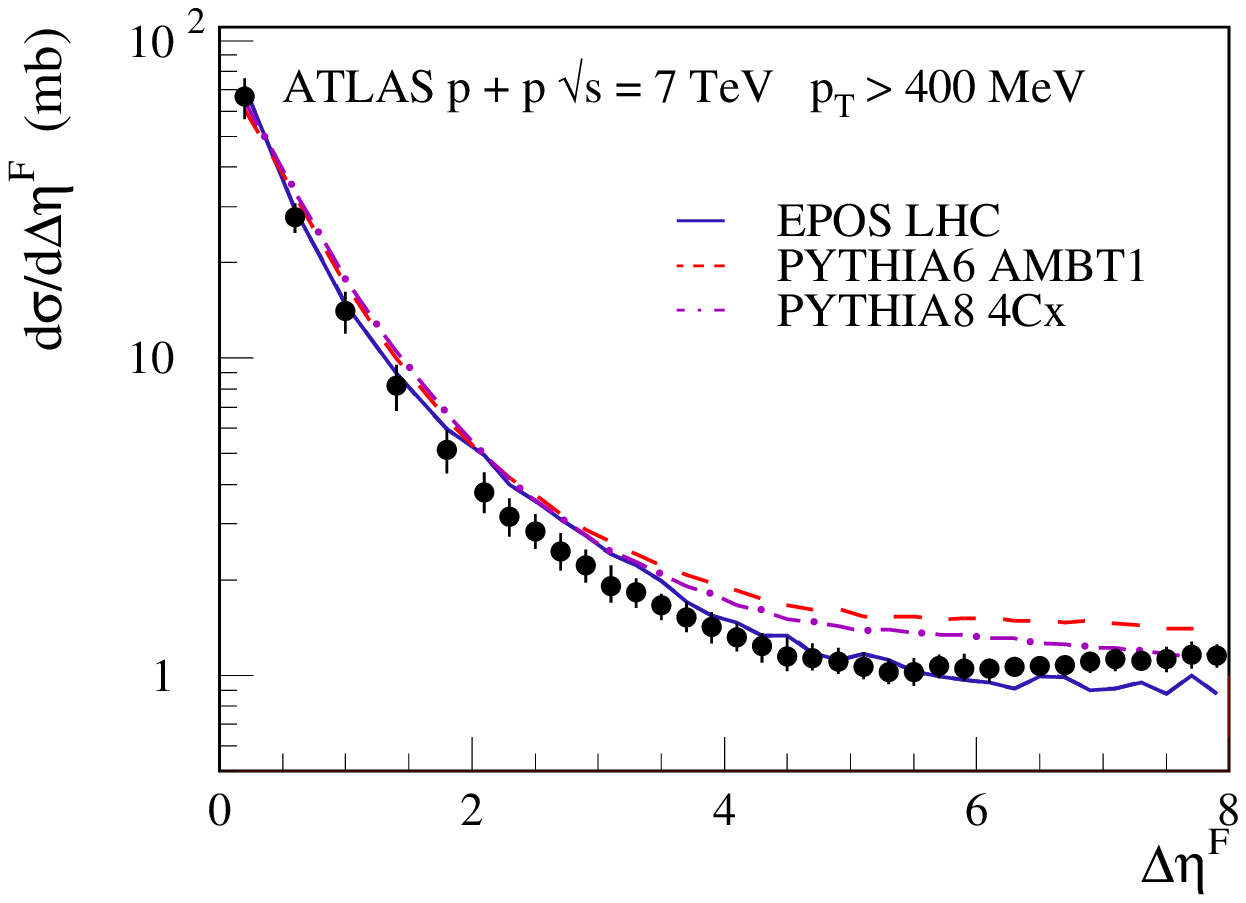}%
\caption{\label{fig_pythia_rapgap}  ATLAS measurement of the pseudorapidity gap  $\Delta \eta F$ for particles with $p_{t,cut} > 400$ MeV in minimum bias events at 7~TeV~\protect\cite{Aad:2012pw}  compared to EPOS~LHC (solid line), Pythia6 (dashed line) and Pythia8 (dash-dotted line) simulations.    }
\end{centering}
\end{minipage}
\end{figure}

The transverse energy flow, which is sensitive to both the charged and neutral charged particle production, at the LHC has also been measured by ATLAS \cite{Aad:2012mfa} over the full acceptance range of the detector as shown in figure~\ref{fig_pythia_sumEtvseta}. While all models predict a too fast decrease of the transverse energy in the forward region compared to the central region, EPOS~LHC describes the data well up 
to  $|\eta|=4.5$. The AMBT1 tune describes the data well in the central region where it was tuned however, only to the charged particles. Apparently the neutral particle production is equally well described by the model. However, in the 
forward region starting at $|\eta| \geq 2.4$, AMBT1 significantly undershoots the data and the disagreement increases towards the forward direction. Tune 4Cx shows a similar $|\eta|$ dependence as AMBT1, however, due to the slightly higher predictions of 
particle production with $p_t \geq 500$ MeV as discussed above, it is slightly high in the central region here and is slightly closer to the data in the forward region. 

The pseudorapidity gap  distribution in minimum bias collisions at 7~TeV has been measured by ATLAS \cite{Aad:2012pw}. The 
cross sections have been measured differentially in terms of $\Delta \eta F$, the larger of the pseudorapidity
regions extending to the limits of the ATLAS acceptance, at $\eta = \pm 4.9$, in which no final state particles
are produced above a transverse momentum threshold
$p_{t,cut}$. At small 
$\Delta \eta F$  the data 
test the reliability of hadronization models in describing rapidity and transverse momentum fluctuations in final state particle production. The measurements at larger gap sizes are dominated by contributions from the single diffractive dissociation process (pp $\rightarrow$ Xp), enhanced by double dissociation (pp $\rightarrow$ XY). Figure~\ref{fig_pythia_rapgap} shows the 
rapidity gap distirbution with a $p_{t,cut} \geq 400$ MeV. All models are describing the small gap region very well, but 
at larger rapidity gaps, where the diffractive processes contribute significantly, is predicted too high by both Pythia tunes. EPOS~LHC, which is tuned to the diffractive cross sections measured at lower energies (SPS) describes this distribution very well. 

\begin{figure}[hptb]
\begin{minipage}[t]{0.48\linewidth}
\begin{centering}
\includegraphics[height=1.\columnwidth,angle=-90]{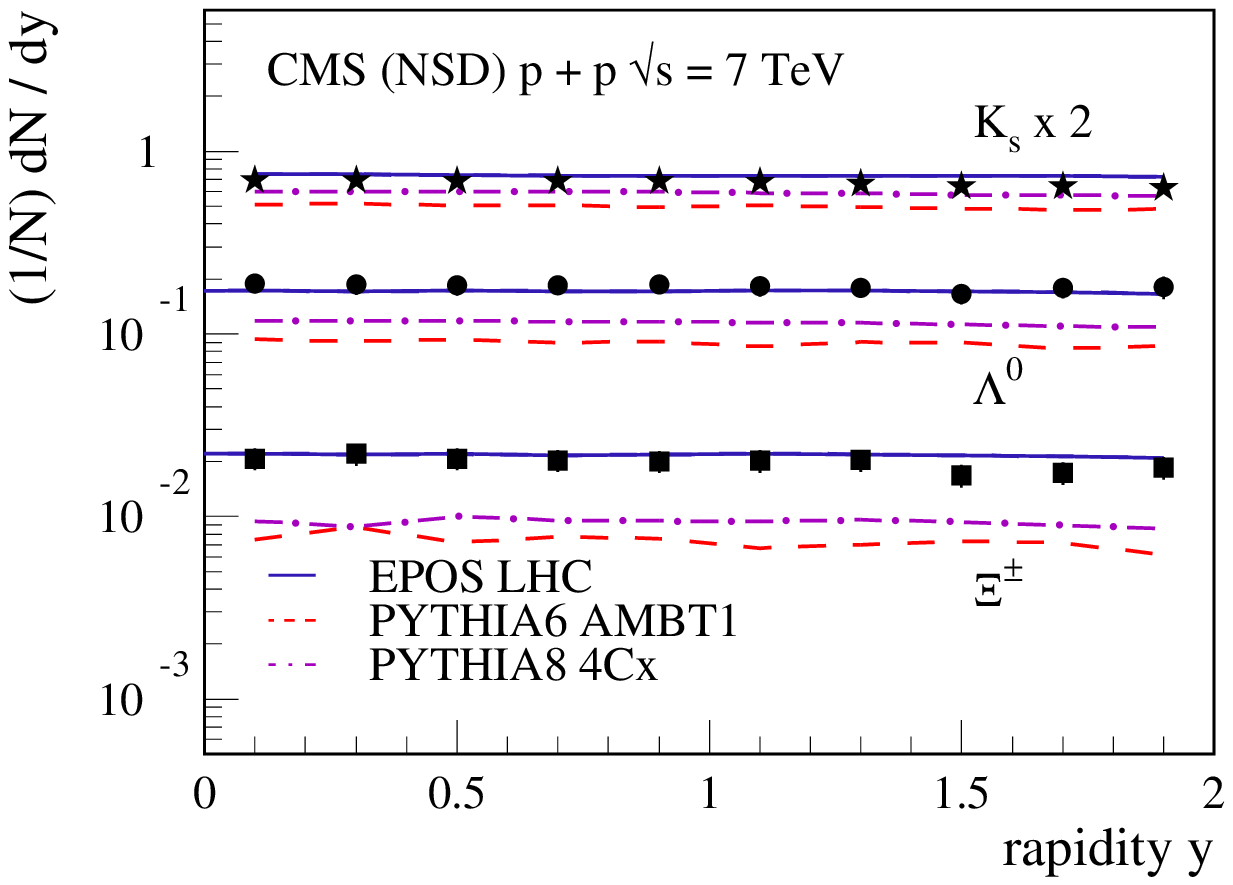}%
\caption{ \label{fig_pythia_id_rate} CMS measurement of strange particles ($K^0_s$, $\lambda^0$ and $\Xi^\pm$) yields in NSD events at 7~TeV~\protect\cite{Khachatryan:2011tm}  compared to EPOS~LHC (solid line), Pythia6 (dashed line) and Pythia8 (dash-dotted line) simulations.}
\end{centering}
\end{minipage}
\hfill
\begin{minipage}[t]{0.48\linewidth}
\begin{centering}
\includegraphics[height=1.\columnwidth,angle=-90]{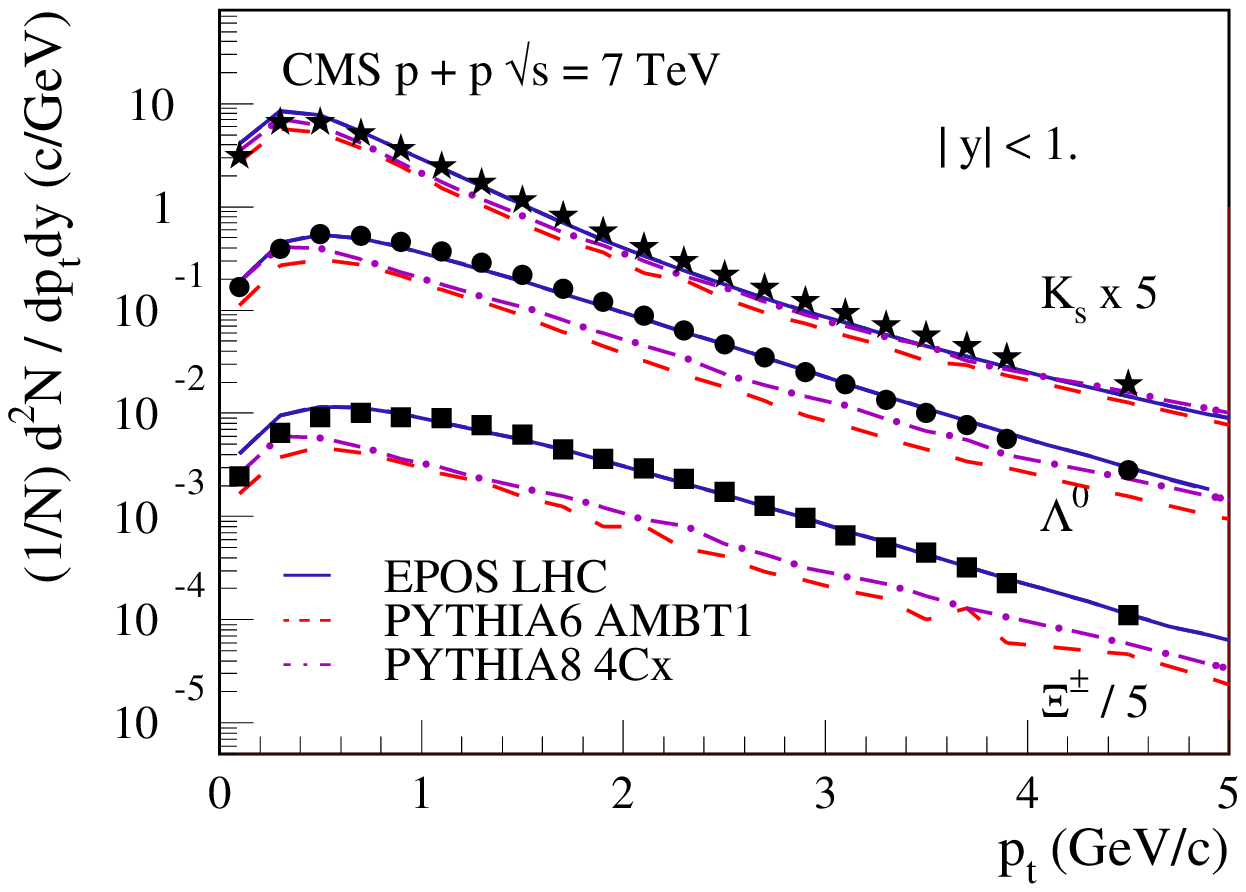}%
\caption{ \label{fig_pythia_id_pt} CMS measurement of transverse momentum distribution of strange particles ($K^0_s$, $\lambda^0$ and $\Xi^\pm$) in NSD events at 7~TeV~\protect\cite{Khachatryan:2011tm}  compared to EPOS~LHC (solid line), Pythia6 (dashed line) and Pythia8 (dash-dotted line) simulations.}
\end{centering}
\end{minipage}
\end{figure}

The EPOS model extensions are also sensitive to the rate and $p_t$ of identified particles~\cite{Aad:2011hd, Chatrchyan:2012qb, Khachatryan:2011tm}, including also particles with strange quarks. Figure~\ref{fig_pythia_id_rate} shows the rate predictions for $\Lambda, \Xi$ and $K_s$ particles as 
measured by CMS~\cite{Khachatryan:2011tm}. Both pythia models describe the  LEP data on fragmentation well, but apparently fail to describe these LHC data. The same problem is observed with ALICE data on multistrange baryon production~\cite{Abelev:2012jp}. EPOS~LHC is able to describe both production (LEP and LHC) over the full rapidity range due to its statistical hadronization effects combined with string fragmentation. The description of the $p_t$ spectra in figure~\ref{fig_pythia_id_pt} is similar: Pythia models show significant deviation while EPOS~LHC reproduce nicely the data which is attribute to its implementation of collective flow effects.

%

\section{Summary}

After a short presentation of the main ingredients of the EPOS~1.99 model 
and in particular the possibility to hadronize part of the secondary
particles including some collective effects, the
new flow parametrization implemented in EPOS~LHC is described. The main change
is that the flow intensity depends only on the total mass of the high density 
core produced by the overlap of string segments due to multiple parton 
interactions (MPI in {\it p-p}) or mutliple nucleon interaction (MNI dominating 
in {\it A-B}). Since the
volume, and as a consequence the speed of the core expansion, is very different
in {\it p-p} and {\it A-B}, this allows two different flow parametrizations to be used for 
the two different systems. In the case of {\it p-A} interactions a smooth transition is
used depending on how the core is created (from MPI or MNI). The core decay
does not follow usual string fragmentation rules but corresponds to a 
statistical decay. In addition to the flow, which will have a strong impact
on transverse momentum distributions, the particle ratios are modified. In 
particular, multistrange baryon formation is favored compared to string
fragmentation. Comparing EPOS~LHC
to various LHC data, it is demonstrated that this approach provides a very good
description of {\it p-p}, {\it Pb-p} and {\it Pb-Pb} data.

A comparison of the EPOS~LHC tune and different Pythia tunes to different LHC measurements is also performed.  Similar good agreement can be reached 
 for the pseudo rapidity distribution of charged particles and the correlation between mean $p_t$ and multiplicity in the 
 minimum bias events. Significant differences are observed in the $p_t$ spectrum 
where the EPOS model describes the data well due to its correlated flow treatment whereas PYTHIA, which lacks such a model, shows 
up to 20\% deviations.  The rate and $p_t$ spectra of the  identified particles is well described by EPOS, while 
significant deviations are observed in PYTHIA. The amount of strange particles is related to the  statistical decay in EPOS which can easily create strange quarks at this rate without changing string hadronization constrained by LEP data.

The core formation, including a transverse flow, is also a key point needed to 
describe in detail even minimum bias {\it p-p} data. These effects can not be 
neglected in particular in {\it Pb-p} scattering where final state interactions 
are even more important than in {\it p-p}. As a consequence any analysis using
particles with $p_t<5$~GeV/c should be interpreted with care since even 
{\it p-p} data may include final state interactions.

\bibliographystyle{JHEP}
\bibliography{references}

\end{document}